\documentclass[prb, amsfonts, twocolumn, a4, showpacs]{revtex4}
\usepackage{graphicx}

\begin{document}

\title[Transport through correlated quantum dots ...]
 {Transport through correlated quantum dots:
 An investigation using the functional renormalization group}

\author{Michael Weyrauch and Dieter Sibold}%\dag}
\affiliation{Physikalisch--Technische Bundesanstalt, D--38116
Braunschweig, Germany}

\date{\today}

\begin{abstract}
Calculations using the (exact) fermionic functional renormalization
group are usually truncated at the second order of the corresponding
hierarchy of coupled ordinary differential equations. We present a
method for the systematic determination of higher order vertex
functions. This method is applied to a study of transport properties
of various correlated quantum dot systems. It is shown that for
large Coulomb correlations higher order vertex functions cannot be
neglected, and a static approximation is insufficient.
\end{abstract}


\email{michael.weyrauch@ptb.de} \maketitle

\section{Introduction}\label{Introduction}

Recently, electron transport through ultrasmall fermionic systems
(quantum dots) has been investigated intensively~\cite{KOEN05,
SIN05, MED06a, KAR06, KAS07}. Motivation for these studies is the
potential application of quantum dots to quantum computers and
quantum measurement devices~\cite{SOH97}. The transport process can
be described using the concept of Coulomb blockade, when Coulomb
correlations are large compared to other energy scales in the
quantum dot system. However, a detailed analysis of the conductance
and level occupancies of the quantum dot states as a function of an
applied gate voltage shows a rich structure, which is related to
Kondo physics.  The theoretical description of such effects requires
non-perturbative methods, and many studies employ Wilson's numerical
renormalization group (NRG). Recently, functional renormalization
group techniques have emerged as a new non-perturbative tool to
describe mesoscopic phenomena theoretically~\cite{MED06a, KAR06}.

While the functional renormalization group equation (fRG) is exact
in principle, for practical applications it must be
truncated~\cite{BER00}.  In particular, most fRG investigations of
mesoscopic systems use a static approximation (i.e.  energy
independent self-energies and higher vertex functions), and the
hierarchy of differential equations corresponding to the fRG is
truncated at second order. Since fRG calculations truncated at
second order are  quite `cheap' numerically in comparison to NRG  or
DMRG (density matrix renormalization group) calculations, it is
possible to map out the potentially huge parameter space of quantum
dot systems with reasonable effort. Functional renormalization group
calculations beyond the static approximation are technically
difficult and require a significant numerical effort as is
demonstrated in model studies for single quantum
particles~\cite{HED04, WEY05}.

The uncertainties incurred by the truncation of the fRG are not
easily controlled. Therefore, it is worth studying the static
approximation beyond second order truncation. Moreover, methods to
go beyond the static approximation with reasonable effort should be
developed, and the limits of the static approximation should be
assessed.

It is the purpose of the present paper to investigate the influence
of vertex functions beyond second order in a static approximation.
To this end, we develop a method which enables to extract the
complete set of ordinary differential equations from the underlying
functional differential equation (section II). Our method uses a
different regularization scheme than the one usually employed for
fermionic systems, which regularizes the free propagator (see e.g.
Refs.~\onlinecite{HED04} and~\onlinecite{SAL99}). Our cutoff
procedure is adapted from the scheme used in
Ref.~\onlinecite{BER00}, which adds a regulator to the action. We
are able to show that the two procedures yield the same set of
ordinary coupled differential equations for the standard hard cutoff
regulator. Furthermore, it is shown that this set is finite in
principle for a fermionic system. (This is in contrast to Bose
systems where this set is infinite.) The number of differential
equations obtained nevertheless grows exponentially with the number
of degrees of freedom of the mesoscopic system under consideration.

For quantum dot systems described by only a small number of
electronic states, a study considering all possible vertex functions
is feasible (within the constraints of the static approximation). In
this way, the limits of the static approximation can be assessed. In
order to do that we investigate transport properties of various
quantum dot devices with different couplings to the external leads
(section III). It is shown that consideration of higher order vertex
functions is important in order to describe transport properties of
correlated quantum dots quantitatively, in particular for systems
with strong Coulomb correlations. However, in many cases we find
that the set of differential equations to be solved develops
singularities if we increase Coulomb correlations beyond some
critical value, and the renormalization flow becomes unphysical.
This is, of course, not unexpected, and just shows that for a
quantitative analysis of mesoscopic systems with strong Coulomb
correlations one eventually needs to go beyond the static
approximation. This will be addressed in a forthcoming publication
using the methods developed here but including a wave function
renormalization in the kinetic energy term of the action.

\section{Functional RG scheme for interacting Fermions}\label{Method}

The effective average action $\Gamma_k$ for interacting Fermions
evolves according to the functional renormalization group equation
\cite{WET93, BER00}
\begin{equation}\label{method-fRG}
\frac{\partial}{\partial k}\Gamma_k [\phi^*,\phi]=-\frac{1}{2}{\rm
Tr}
    \left\{ [{\Gamma}_k^{(2)}[\phi^*,\phi]+R_k]^{-1}\frac{\partial {R}_k}{\partial
    k}\right\}.
\end{equation}
Here, we assume that the space of possible states of the Fermions is
suitably discretized, so that a $N$-component vector of Grassmann
variables $\phi(\tau)=(\phi_1(\tau),\ldots,\phi_N(\tau))$ describes
the evolution of the interacting Fermion system in imaginary time
$\tau$. Each index $\alpha\in\{1,\ldots, N\}$ represents all quantum
numbers necessary to completely specify a Fermionic state, e.g. spin
projection $\sigma$ and position $j$ in a linear chain of electrons.
The functional derivative of $\Gamma_k$ with respect to $\phi^*$ and
$\phi$ is denoted by $\Gamma_k^{(2)}$. The regulator ${R}_k$ is
introduced in order to suppress thermal and quantum fluctuations at
energy or momentum scales $k$ larger than any physical scale
relevant for the problem that is beeing investigated. With
decreasing $k$, the regulator gradually ``switches on" such
fluctuations until they are fully included at $k=0$, i.e. at $k=0$
the regulator $R_k$ vanishes. A concrete choice for $R_k$ will be
discussed below.  The initial condition for the evolution described
by Eq.~(\ref{method-fRG}) is obtained from the Hamiltonian defining
the problem to be solved~\cite{BER00}. The trace in
Eq.~(\ref{method-fRG}) is to be performed over all relevant quantum
numbers.

In order to  solve a functional differential equation like
Eq.~(\ref{method-fRG})  in practice, it must be truncated. This
essentially entails a suitable transformation of the functional
differential equation into an infinite set of coupled partial or
ordinary differential equations. This set of differential equations
is then truncated at a suitable order. The standard technique
expands both sides of the equation for the effective average into a
Taylor series about a $\tau$ independent vector $\phi_0$ using
$\phi(\tau)=\phi_0+\psi(\tau)$. One then obtains an infinite set of
ordinary or partial differential equations for the  Taylor
coefficients of the various powers in $\psi(\tau)$.

For most practical applications, the truncation procedure described
so far is still too general, and we need further simplifications:
One standard way to proceed is to prescribe the functional form of
the effective average action more specifically: In this paper we
assume that the effective average action takes the form
\begin{equation}\label{method-Gamma}
\Gamma_k[\phi^*,\phi]=\int_0^\beta {\rm d}\tau
\sum_{\alpha=1}^N\phi_\alpha^*(\tau)\frac{\partial}{\partial\tau}\phi_\alpha(\tau)+
       U_k\left(\phi^*(\tau),\phi(\tau)\right)
\end{equation}
where the `effective potential' $U_k$ does not depend on derivatives
of the Grassmann variables with respect to the imaginary time
$\tau$. Assuming the specific form (\ref{method-Gamma}) yields
energy independent (static) self-energies and higher order vertex
functions.  Of course, more general functional forms for $\Gamma_k$
lead to energy dependent vertex functions [e.g. a wave function
renormalization $Z_k\left(\phi^*(\tau),\phi(\tau)\right)$
multiplying the kinetic energy term in Eq.~(\ref{method-Gamma})],
but such a truncation will not be investigated in this paper.

Since the effective potential $U_k$ must be a Grassmann scalar, its
form in terms of $\phi(\tau)$ is fairly well determined: It must be
a linear combination of all possible products of elements of the set
of Grassmann variables $\{\phi_\alpha^*, \phi_\alpha\}$ with an even
number of factors. Due to the nil-potency of the Grassmann variables
the number of possible products is finite, i.e. the number of terms
in the linear combinations is finite. Since the number of particles
is conserved in the physical models that we want to investigate, we
only need to include terms with an equal number of $\phi_\alpha^*$
and $\phi_\alpha$, which correspond to Fermion creation and
destruction, respectively.

Here is an example for $N=2$: Due to the nil-potency of the
Grassmann variables, the effective potential has the general form
\begin{eqnarray}\label{method-Uform}
 U_k(\phi^*,\phi) &=& a_{0,k} + \phi^*_{1}\phi_{1}a_{11,k} +
\phi^*_{2}\phi_{2}a_{22,k}+ \\
& &\phi^*_{1}\phi_{2}a_{12,k}+\phi^*_{2}\phi_{1}a_{21,k}+
\phi^*_{1}\phi^*_{2}\phi_{1}\phi_{2} a_{1212,k}.\nonumber
\end{eqnarray}
The $k$-dependent coefficients in this expression will be called
`running couplings'. The running couplings  have direct physical
significance: Since the effective average action at $k=0$ is the
generator of the vertex functions, each running coupling corresponds
to an $n$-point vertex function, where $n$ corresponds to its number
of Grassmann factors. In particular, $a_0$ is directly related to
the ground state energy of the Fermion system, and the $a_{ij,k}$
correspond to the self-energies. For simplicity, the $k$-dependence
of the couplings will not always be explicitly indicated.

Following the general procedure outlined above, we will now
determine flow equations for the running couplings. To this end we
first expand the effective average action~(\ref{method-Gamma}) about
$\tau$-independent Grassmann variables
$\phi_{0}=(\phi_{01},\ldots,\phi_{0N})$, i.e. $\phi_{\alpha}(\tau) =
\phi_{0\alpha}+\psi_{\alpha}(\tau)$ with
$\phi_{0\alpha}=\phi_\alpha(0)$. Up to second order in
$\psi_\alpha(\tau)$, the expansion is given by
\begin{eqnarray}\label{method-Gamma-exp}
\Gamma_k[\phi^*,\phi]&=&\beta U^{(0)}_{k}(\phi^*_0,\phi_0)\nonumber\\
& &+\frac{1}{2}\int_0^\beta {\rm d}\tau \int_0^\beta {\rm d}\tau^\prime
\Psi^\dagger(\tau)\gamma_k^{(2)}(\phi_0^*,\phi_0)\Psi(\tau^\prime)\nonumber\\
& &+\ldots
\end{eqnarray}
with
$\Psi(\tau)=({\psi_{1}}(\tau),\ldots,{\psi_{N}}(\tau),{\psi_{1}^*}(\tau),\ldots{\psi_{N}}^*(\tau))$.
The lowest order term is determined by the effective potential
\begin{equation}
U^{(0)}_{k}(\phi^*_0,\phi_0)=\left.U_k(\phi^*(\tau),\phi(\tau))\right|_{\psi^*=\psi=0}.
\end{equation}
The second order term contains the second derivative of the effective potential
\begin{eqnarray}
\gamma_k^{(2)}(\phi_0^*,\phi_0)&=&\left(\mathbb{E}^\prime\partial_\tau+U_{k}^{(2)}(\phi_0^*,\phi_0)\right)\delta(\tau-\tau^\prime),\nonumber\\
\quad U_{k,\alpha\beta}^{(2)}(\phi^*_0,\phi_0)&=& \left.\frac{\partial^2 U_k}{\partial \Psi_\alpha \partial\Psi_\beta^\dagger }\right|_{\psi^*=\psi=0}
\end{eqnarray} with
 $\mathbb{E}^\prime={\rm diag}(1,\ldots,1,-1,\ldots,-1)$. The
term in first order in $\Psi$ drops out. For example, for $N=2$
assuming the symmetry $a_{12}=a_{21}$, one easily finds from
Eq.~(\ref{method-Uform}) the matrix
\begin{widetext}
\begin{equation}\label{method-U}
U_{k}^{(2)}= \left(
\begin{array}{cccc}
a_{11}-\phi_{02}^*\phi_{02}a_{1212} & a_{12}+{\phi^*_{02}} {\phi_{01}} a_{1212} & 0 & {\phi_{01}} {\phi_{02}} a_{1212} \\
a_{12}+{\phi^*_{01}} {\phi_{02}}a_{1212} & a_{22}-{\phi^*_{01}} {\phi_{01}}a_{1212} & -{\phi_{01}}{\phi_{02}}  a_{1212} & 0 \\
0 & {\phi_{01}^*} {\phi_{02}^*} a_{1212} & -a_{11}+\phi_{02}^*\phi_{02}a_{1212} & -a_{12}-{\phi_{01}^*}{\phi_{02}} a_{1212} \\
-{\phi_{01}^*}{\phi_{02}^*}  a_{1212} & 0 & -a_{12}
-{\phi_{02}^*}{\phi_{01}} a_{1212}&
-a_{22}+\phi^*_{01}\phi_{01}a_{1212}
\end{array} \right)
\end{equation}
\end{widetext}

In order to do the $\tau$-integration in Eq.~(\ref{method-Gamma-exp}) we
Fourier transform the Grassmann variables $\psi_\alpha$ using
\begin{eqnarray}
\psi_\alpha(\tau)=\frac{1}{\sqrt{\beta}}\sum_n e^{i\omega_n\tau}\psi_{\alpha,n}
\end{eqnarray}
with the fermionic Matsubara frequency $\omega_n=(2n+1)\pi/\beta$
and obtain for the effective average action
\begin{equation}\label{method-expGamma}
\Gamma_k[\phi^*,\phi]
       = \beta U^{(0)}_{k}+\frac{1}{2}\sum_{n=-\infty}^\infty \Psi^\dagger_n (i\omega_n\mathbb{E}+U_k^{(2)})\Psi_n.
\end{equation}
Here $\mathbb{E}$ denotes the unit matrix.
Moreover, we assume a regulator  which is diagonal in
Matsubara frequency space
\begin{equation}\label{method-Rk}
\sum_n \psi^*_n R_{k,n} \psi_n=\frac{1}{2}\sum_n\Psi^\dagger_n
R_{k,n}\mathbb{E}^\prime\Psi_n.
\end{equation}

Inserting $\Gamma_k$ and $R_k$ into Eq.~(\ref{method-fRG}) and
comparing terms up to order zero in $\psi$ we obtain the following
equation
\begin{equation}\label{G-renorm}
\frac{\partial}{\partial k} U^{(0)}_{k}
  =-\frac{1}{2\beta}\sum_n{\rm
Tr}
  K^{-1}
  \frac{\partial}{\partial k}{R}_{k,n}.
\end{equation}
with $K=(i\omega_n+U_k^{(2)}){\mathbb{E}^\prime}+R_{k,n}\mathbb{E}$.
From this equation we extract a set of ordinary differential
equations for the various coupling constants by comparing the
coefficients of the various monomials of Grassmann variables. In
order to do this, it is necessary to invert the Grassmann matrix
$K$. The matrix is inverted using the formula
\begin{equation}\label{method-inversion formula}
K^{-1}=(K_0+K_1)^{-1}=K_0^{-1}\sum_{m=0}^N (-K_1K_0^{-1})^m
\end{equation}
where $K_0$ is chosen to be the body of $K$ (Grassmann scalar part
of $K$) and $K_1$ its soul (Grassmann non-scalar part). That the sum
in Eq.~(\ref{method-inversion formula}) only runs up to $N$ is due
to the fact that each matrix element of $K_1$ is bilinear in the
Grassmann variables, so that after $N$ factors of $K_1$ the sum must
terminate. Obviously, the inverse of $K$ does not exist if the
determinant of the body of $K$ vanishes. In this case, the
renormalization scheme is not well defined.

At $T=0$ the sum over the Matsubara frequencies in Eq.~(\ref{G-renorm}) converts into an integral
\begin{equation}
\frac{1}{\beta}\sum_n \rightarrow \int_{-\infty}^\infty \frac{{\rm
d}\omega}{2\pi}.
\end{equation}
This integral is most easily evaluated for the sharp cut-off regulator
\begin{equation}\label{method-regulator}
R_k(\omega)=Ck\theta(k^2-\omega^2)
\end{equation}
with $C$ a suitably chosen large constant. This regulator fulfills
the general requirements that it vanishes for $k=0$ and it dominates
the effective average action for $k\rightarrow\infty$. Furthermore,
it has the advantage that the integration over the Matsubara
frequencies can be done analytically. Some technical issues related
to this integration will be discussed in Appendix A.

An often used approach  to the fermionic functional renormalization
group  applies a hard cutoff in such a way that the free propagator
at large $k$ is suppressed, and propagation is gradually `switched
on' when $k$ is reduced (see e.g. Refs.~\onlinecite{SAL99}
and~\onlinecite{HED04}). In the following we shall see that this
(standard) method and the cutoff employed here yield identical sets
of flow equations.

With the results from Appendix A, we can write Eq.~(\ref{G-renorm})
for the hard cut-off regulator (\ref{method-regulator}) in the
following form
\begin{eqnarray}
\frac{\rm d}{{\rm d}k}a_{0,k}&=&\frac{1}{4\pi} \sum_{\lambda=\pm k}\log {\rm det}
\left(\frac{1}{k}G^{-1}_k(i\lambda)\right),\label{method-floweqa}\\
\frac{\rm d}{{\rm d} k} (U^{(0)}_{k}-a_{0,k})
&=&-\frac{1}{4\pi}\sum_{m=1}^N\sum_{\lambda=\pm k}\frac{1}{m}{\rm
Tr}\left(-M_kG_k(i\lambda)\right)^m. \label{method-floweqb}\nonumber\\
\end{eqnarray}
Eq.~(\ref{method-floweqb}) represents a {\it finite} set of coupled ordinary differential equations for the running couplings.
The  block-diagonal $2N\times 2N$ matrix  $G_k(i\lambda)$
is given by
\begin{equation}\label{method-propagator}
  G_k(i\lambda)=\left(\begin{array}{cc}g_k(i\lambda) &0\\0&g_k(-i\lambda)\end{array}\right)=
  \left(\begin{array}{cc}b_k(i\lambda) &0\\0&b_k(-i\lambda)\end{array}\right)^{-1},
\end{equation}
and the matrix elements of the $N\times N$ submatrix $b_k(i\lambda)$
are essentially determined by the self-energies
\begin{eqnarray}\label{method-b}
  b_{\alpha\beta,k}(i\lambda)=a_{\alpha\beta,k}+i\lambda\delta_{\alpha\beta}.
\end{eqnarray}
The $2N\times 2N$ matrix $M_k$ represents the soul of
$U_k^{(2)}\mathbb{E}^\prime$ and contains all running couplings
except the self-energies.

Eqs.~(\ref{method-floweqa}) and (\ref{method-floweqb}) are the main
results of this section. From these results, the flow equations for
the various running couplings can be directly extracted up to the
desired order by comparing the coefficients of the different
monomials of Grassmann variables. As will be discussed further
below, up to second order in $m$, Eqs.~(\ref{method-floweqa}) and
(\ref{method-floweqb}) are equivalent to results obtained in the
literature using a hard cutoff on the free propagator. The advantage
of Eqs.~(\ref{method-floweqa}) and (\ref{method-floweqb}) is that
they allow an easy and systematic construction of higher order
contributions.

In order to illustrate the use of Eqs.~(\ref{method-floweqa}) and
(\ref{method-floweqb}), we extract the flow equations for $N=2$
assuming the symmetry  $a_{12}=a_{21}$ and purely real running
couplings. In this case, the submatrix $g_k(i\lambda)$ is given by
\begin{eqnarray}
  g_k(i\lambda)&=&\frac{1}{b_{11}b_{22}-b_{12}^2}
  \left(\begin{array}{cc}b_{22}&-b_{12}\\-b_{12}&b_{11}\end{array}\right)
\end{eqnarray}
with $b_{\alpha\beta}$  defined in Eq.~(\ref{method-b}). The matrix
$M_k$ can be directly read off from Eq.~(\ref{method-U})
\begin{equation}
M_k=a_{1212,k}\left(
\begin{array}{cccc}
-\phi_{02}^*\phi_{02} & {\phi^*_{02}} {\phi_{01}}  & 0 & {\phi_{01}} {\phi_{02}}  \\
{\phi^*_{01}} {\phi_{02}} & -{\phi^*_{01}} {\phi_{01}} & -{\phi_{01}}{\phi_{02}}   & 0 \\
0 & -{\phi_{01}^*} {\phi_{02}^*}  & -\phi_{02}^*\phi_{02} & {\phi_{01}^*}{\phi_{02}}  \\
{\phi_{01}^*}{\phi_{02}^*}  & 0 &{\phi_{02}^*}{\phi_{01}} &-\phi^*_{01}\phi_{01}
\end{array} \right).
\end{equation}
After some algebra
one obtains the following flow equations for the various
running couplings
\begin{eqnarray}\label{method-2com-result}
a_0'&=&\frac{1}{ \pi } {\rm Re}\log \left(b_{11} b_{22}-b_{12}^2\right)/k^2,\nonumber\\
a_{ij}'&=&-\frac{a_{1212}}{\pi}{\rm Re}\frac{ b_{ij}}{b_{11}b_{22}-b_{12}^2},\\
a_{1212}'&=&-\frac{a_{1212}^2}{\pi}{\rm Re}\frac{1}
{b_{11} b_{22}-b_{12}^2}\nonumber\\
& &-\frac{a_{1212}^2}{2\pi}\frac{b_{11}^* b_{22}+b_{11}
b_{22}^*-2 b_{12} b_{12}^* }{\left(b_{11} b_{22}-b_{12}^2\right)
\left(b_{11}^*
b_{22}^*-b_{12}^{*2}\right)}.\nonumber
\end{eqnarray}
The prime indicates a derivative with respect to $k$, and we do not
indicate explicitly the $k$ dependence of the running couplings.

For $N=2$ the result~Eq.~(\ref{method-2com-result}) is complete
within the static approximation. This means, that due to the finite
number of Grassmann monomials which can be constructed from four
Grassmann variables, there will be no higher order equations. The
right hand side of the equations for $a_{ij}$ arises from $m=1$ in
the sum over $m$ in Eq.~(\ref{method-floweqb}), while the right hand
side of the equation for $a_{1212}$ is the contribution from $m=2$.
Eqs.~(\ref{method-2com-result})
 are equivalent to Eqs.~(26) and~(27) in
Ref.~\onlinecite{KAR06}, where they are used to study electron
transport through a single quantum dot. Moreover, the equations for
$a_{ij}$ in Eq.~(\ref{method-2com-result}) are equivalent to Eq. (1)
in Ref.~\onlinecite{MED06a} which was used to show that there are
unusual correlation induced resonances in a polarized double dot
system. (In order to show the equivalence one must set $a_{1212,k}$
to its initial condition $a_{1212,k}=U$.)

We now discuss the structure of the set of differential equations
obtained from Eq.~(\ref{method-floweqb}) for larger $N$ using $N=4$
as an example. As can be easily established, the number of
differential equations grows exponentially. In particular, for $N=4$
we find 35 coupled equations corresponding to the number of
Grassmann monomials we can build from 8 Grassmann variables in a
system with particle conservation. If symmetries between these
Grassmann monomials are considered the number of equations reduces,
e.g. for a system of four electrons with  spin symmetric
interactions we only obtain 26 independent equations. It turns out
that the resulting (complete) set of equations is quite formidable
and can only be generated with the help of computer assisted
symbolic computation. Nevertheless, some features of this set of
equations can be established in general. These follow from the
structure of Eq.~(\ref{method-floweqb}): A term of order $m$ in the
sum over $m$ in Eq.~(\ref{method-floweqb}) only contributes to
equations for coefficients with at least $2m$ indices. Such a term
contains $m$ propagators $G_k$. E.g., for $N=4$, the term $m=4$ in
the sum over $m$ only contributes to the equation for the running
coupling $a_{12341234}$. Of course, the equation for this particular
running coupling also receives contributions from $m=1, 2, 3$.

To further illustrate the use of Eq.~(\ref{method-floweqb}), we
derive equations suitable for the description of a one dimensional
chain of nearest neighbor coupled spinless electrons (see e.g.,
Ref.~\onlinecite{Med03}). To this end we first need to write down
the effective potential in terms of the running couplings
\begin{eqnarray}
 U(\phi^*,\phi)&=&\sum_{j=1}^N a_{jj}\phi_j^*\phi_j+a_{j,j+1}\phi_j^*\phi_{j+1}+
a_{j,j-1}\phi_{j}^*\phi_{j-1}\nonumber\\
& &+U\sum_{j=1}^N\phi^*_j\phi_j\phi^*_{j+1}\phi_{j+1}
\end{eqnarray}
with periodic boundary conditions $\phi_0=\phi_N$. Here, for
simplicity, it is assumed that the density-density interaction
strength $U$ does not renormalize and remains at its initial value.
In order to obtain the flow equations for the self-energies
$a_{jj^\prime}$, we consider the term $m=1$ in the sum over $m$ in
Eq.~(\ref{method-floweqb}). One easily finds
\begin{eqnarray}
{\rm Tr} M_k G_k&=& U\sum_{j=1}^N
\left[g_{j-1,j-1}(i\lambda)+g_{j+1,j+1}(i\lambda)\right]\phi^*_{0j}\phi_{0j}\nonumber\\
                            &&-\left[g_{j,j+1}(i\lambda)\phi^*_{0j}\phi_{0j+1}+
                            g_{j,j-1}(i\lambda)\phi^*_{0j}\phi_{0j-1}\right]\nonumber\\
                            && +~(i\lambda\leftrightarrow-i\lambda).
\end{eqnarray}
Inserting this result into Eq.~(\ref{method-floweqb}) and comparing
corresponding terms yields the flow equations
\begin{eqnarray}
a_{jj}^\prime&=&\frac{U}{2\pi}\sum_{\lambda=\pm k}\left(g_{j+1,j+1}(i\lambda)+g_{j-1,j-1}(i\lambda)\right)\nonumber\\
a_{j,j\pm 1}^\prime&=&-\frac{U}{2\pi}\sum_{\lambda=\pm k}g_{j,j\pm 1}(i\lambda)
\end{eqnarray}
This result agrees with the one obtained using a hard cutoff on the
free propagator; it is used to study e.g. persistent currents in
mesoscopic rings in Ref.~\onlinecite{Med03}.

\section{Transport through correlated quantum dots}

In the following we study transport through various quantum dot (QD)
systems using the ``effective potential" (i.e. static) approximation
introduced in the previous section. We will emphasize the role of
higher order contributions beyond $m=2$ in the sum over $m$ in
Eq.~(\ref{method-floweqb}). Up to $m=2$, the equations resulting
from Eq.~(\ref{method-floweqb}) can be shown to be equivalent to the
flow equations employed in Ref.~\onlinecite{KAR06}. In order to
elucidate the role of higher order contributions,  we will  study
similar parameter sets as considered in Ref.~\onlinecite{KAR06}. In
some cases, calculations using Wilson's numerical renormalization
group are available for comparison.

\begin{figure}
\unitlength1cm
\begin{picture}(18,3)(0,0)
 \put(2,0)    {\includegraphics[width=5cm]{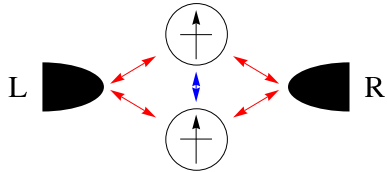}}
\end{picture}
\caption{\label{polarizeddoubledot}\it  (color online) Polarized double quantum
dot coupled to left and right leads.  The arrows between the dots and the leads symbolize the
tunnel couplings $t_j^l$. The arrow between the dots symbolizes interdot Coulomb interactions.}
\end{figure}

The standard model Hamiltonian for the description of QD systems
(depicted schematically in Fig.~\ref{polarizeddoubledot}) consists
of three essential terms: the dot Hamiltonian $H_{\rm D}$, the
Hamiltonian of the non-interacting electrons of the leads $H_{\rm
E}$ and the interaction $H_{\rm DE}$ between the QD and the leads,
\begin{equation}
H=H_{\rm D}+H_{\rm E}+H_{\rm DE}.
\end{equation}
The dot Hamiltonian $H_{\rm D}$ is given by three terms
\begin{eqnarray}
H_{\rm D}&=&\sum_{j\sigma}(\epsilon_{j\sigma}+V_g)d^\dagger_{j\sigma}d_{j\sigma}\nonumber\\
   & &-   \sum_{j>j^\prime,\sigma}t_{jj^\prime}d^\dagger_{j\sigma}d_{j^\prime\sigma}+h.c.\nonumber\\
   & & +\frac{1}{2}\sum_{jj^\prime, \sigma\sigma^\prime}U^{\sigma\sigma^\prime}_{j j^\prime}n_{j\sigma}n_{j^\prime\sigma^\prime}
\end{eqnarray}
where $d^\dagger_{j\sigma}$ ($ d_{j\sigma}$) creates (annihilates)
an electron in state $j$ with spin $\sigma$, respectively. The site
occupancy operator is denoted by
$n_{j\sigma}=d^\dagger_{j\sigma}d_{j\sigma}$.
The  dot levels $\epsilon_{j\sigma}$ may be shifted by application of a
gate voltage $V_g$. The matrix elements $t_{jj^\prime}$ represent
hopping amplitudes between dot levels and $U^{\sigma\sigma^\prime}_{j j^\prime}$ describes
electron-electron interactions. Of course, $U^{\sigma\sigma}_{j j}=0$.

The non-interacting lead electrons created (annihilated) by $c^\dagger_{j\sigma l}$
($c_{j\sigma l}$) at the Fermi surface of the right lead ($l=R$) and left lead  ($l=L$)  are
described by the Hamiltonian
\begin{equation}
H_{\rm E}=-\tau_h\sum_{ j \sigma l} c^\dagger_{j\sigma l}c_{(j+1)\sigma l }+h.c.
\end{equation}
with a hopping amplitude $\tau_h$ between neighboring sites.
The coupling of the leads to the quantum dot levels is modeled by a tunneling Hamiltonian
\begin{equation}
H_{\rm DE}=-\sum_{j\sigma l} t_j^l c^\dagger_{0,\sigma,l} d_{j, \sigma} + h.c.
\end{equation}
where electrons can tunnel from or to the dot levels from site $0$
of the right or left lead. The tunneling matrix elements $t_j^l$ are
assumed to be real in the present paper. They determine the
broadening of the dot levels as will be discussed in more detail
below.
\begin{figure*}
\unitlength1cm
\begin{picture}(18,5.5)(0,0)
 \put(0.,0.0) {\includegraphics[width=4.8cm]{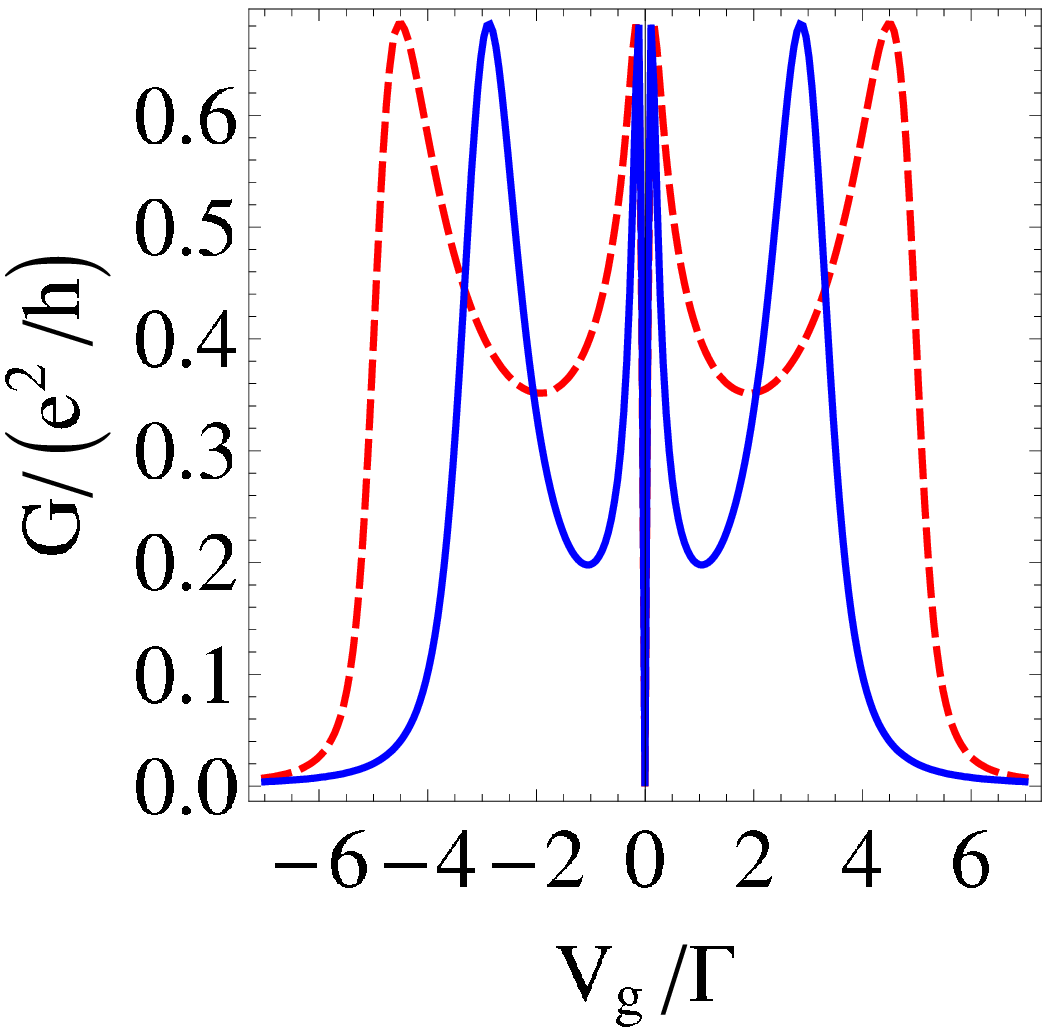}}
 \put(4.2, 4.2) {(a)}
 \put(5.5,0)  {\includegraphics[width=4.8cm]{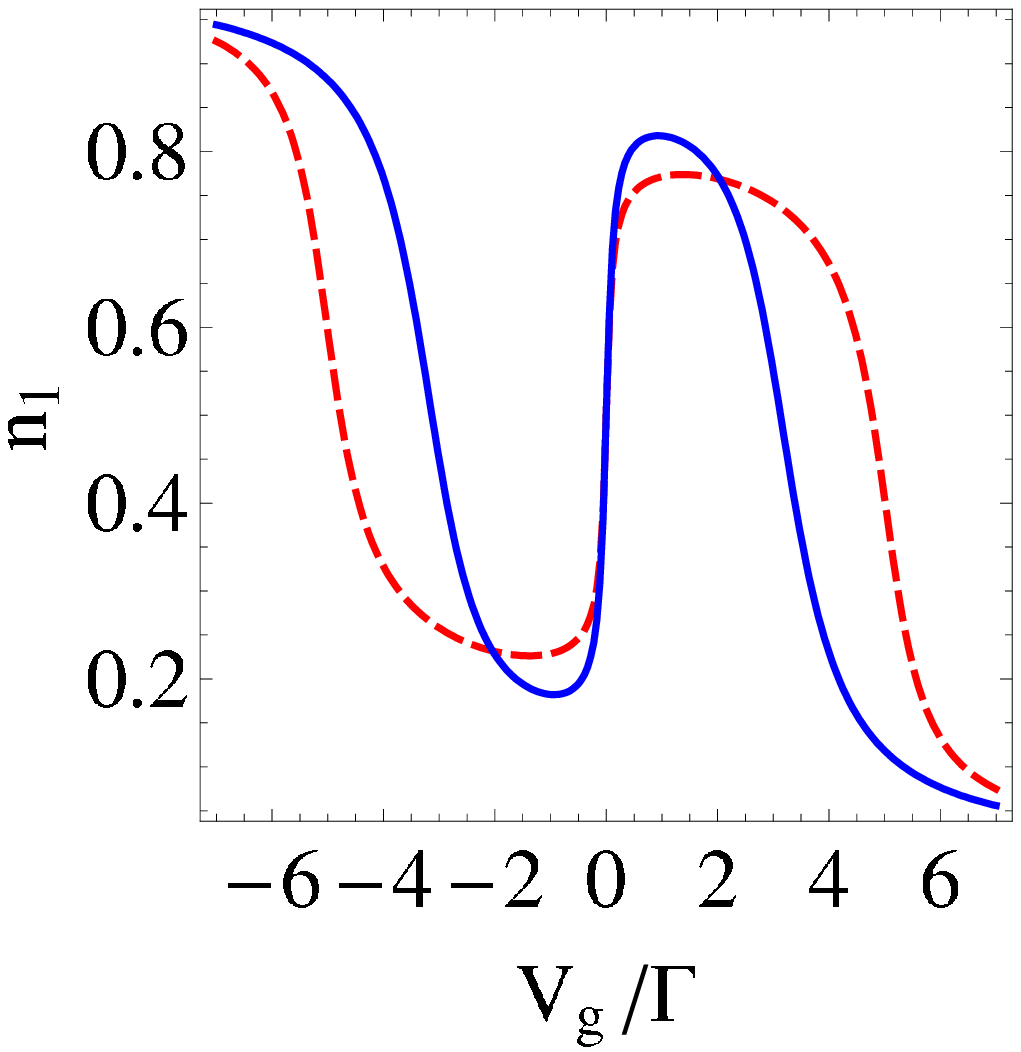}}
 \put(9.5, 4.2) {(b)}
 \put(11.,0.0){\includegraphics[width=4.8cm]{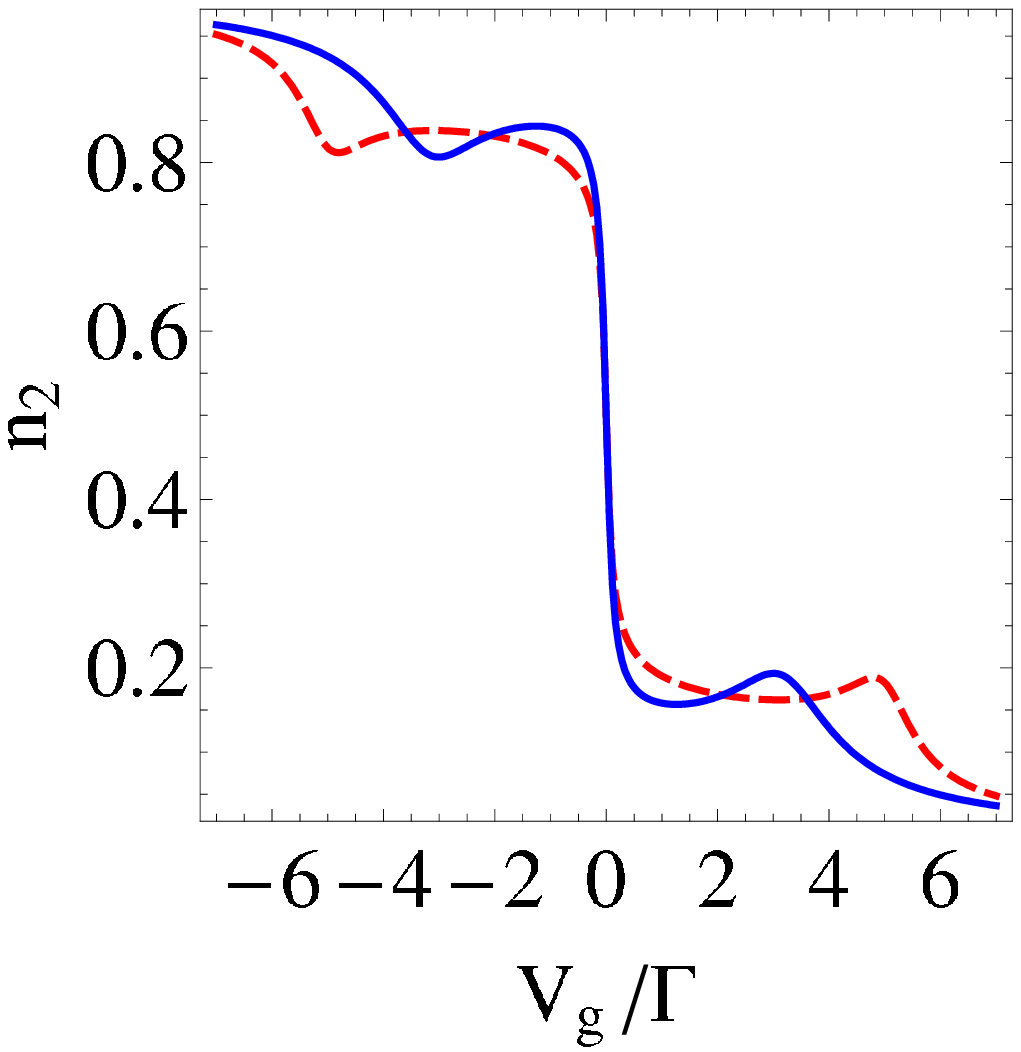}}
 \put(14.8, 4.2) {(c)}
\end{picture}
\caption{\label{den-pol-double-dot}\it (color online) Conductance
$\mathcal{G}$ (a) and occupancies $n_1$ (b) and $n_2$ (c) for a
polarized double dot, $U/\Gamma=8$, $\Gamma_1^L=0.27\Gamma$,
$\Gamma_1^R=0.33\Gamma$, $\Gamma_2^L=0.16\Gamma$,
$\Gamma_2^R=0.24\Gamma$, $t_2^R<0$ all other $t_i^l$ positive,
dashed lines: truncation at $m=1$, full  lines: truncation at
$m=2$.}
\end{figure*}

The Hamiltonian $H$ is translated into an action suitable for a
coherent state path integral formulation (see e.g.,
Ref.~\onlinecite{NEG98}) essentially by replacing creation and
annihilation operators by corresponding Grassmann variables.
Physical observables  at temperature $T=0$ are obtained from the
renormalized Green's functions of the correlated dot system (see
Eq.~(\ref{method-propagator})),
\begin{equation}
g_{ij}^{-1}(i\omega)=i\omega\delta_{ij}+a_{ij,k=0}.
\end{equation}
In a static approximation, the $a_{ij,k=0}$ do not depend on
$\omega$. The dot Green's functions $g$ are $N\times N$ matrices
where $N$ is the total number of  electronic states included in the
description of the quantum dot system. The matrix elements of the
dot Hamiltonian $\epsilon_{j \sigma}+V_g$, $t_{jj^\prime}$, and
$U^{\sigma\sigma^\prime}_{j j^\prime}$ provide the necessary initial
values for the corresponding flow equations. All other flow
equations start the flow from an initial value of zero.

Since the dot system is coupled to the leads via the tunneling
Hamiltonian $H_{\rm DE}$, the dot levels acquire an imaginary part
which is determined using a standard projection  of the full
Hamiltonian on the dot Hamiltonian as described in detail e.g. in
Refs.~\onlinecite{HEW93} and~\onlinecite{ENS05},
\begin{equation}
\tilde{g}_{ij}^{-1}(i\omega)=i\omega\delta_{ij}+a_{ij,k=0}-i{\rm sgn}(\omega)\Delta_{ij}
\end{equation}
with the imaginary part given by
\begin{equation}
\Delta_{ij}=\pi\rho_0\sum_{l=L,R}t_i^{l} t_j^{l}
\end{equation}
in terms of the density of states of the lead electrons at the Fermi level $\rho_0\approx 1/(\pi\tau_h)$
(in the wide band approximation).
Usually, the absolute values of the tunneling matrix elements are given
in terms of the parameters $\Gamma_i^l=\pi\rho_0(t_i^l)^2$.

From this propagator, the average dot occupancies $\langle
n_{j\sigma}\rangle$ at $T=0$  may be easily calculated using
\begin{equation}
\langle n_{j\sigma}\rangle=\frac{1}{2\pi} \int_{-\infty}^\infty {\rm d\omega} e^{i\omega0^+} \tilde{g}(i\omega).
\end{equation}
where the exponential factor enforces convergence of the integral at
large $\omega$. To relate the dot propagator to the conductance is
quite an intricate problem which has been addressed using both the
Landauer-B{\"u}ttiker~\cite{BRU04} and the Kubo
formalism~\cite{ENS05a}. At $T=0$, one obtains for the conductance
$\mathcal{G}$ the rather intuitive result
\begin{equation}\label{dot-conductance}
\mathcal{G}=4\frac{e^2}{h}\left|\sum_{i,j}\pi\rho_0 t_i^{R} t_j^{L}\tilde{g}_{ij}(0)\right|^2
\end{equation}
where the sum runs over all dot levels connected to the leads via
non-vanishing tunneling matrix elements $t^l_{j}$. The phase of the
sum in Eq.~(\ref{dot-conductance}) is known as the  transmission
phase. This quantity is accessible experimentally and will be also
briefly discussed in the following.

\subsection{Polarized double dot}\label{polDD}

We first review calculations for a polarized double dot
corresponding to two states with
$\epsilon_{1\sigma}=\epsilon_{2\sigma}=0$. Each state is occupied by
only one electron as depicted in Fig.~\ref{polarizeddoubledot}. The
system is spin polarized, which may be achieved by putting it into a
strong magnetic field. Studies of this system using different
methods were presented e.g., by K{\"o}nig and Gefen~\cite{KOEN05},
Sindel {\it et al.}~\cite{SIN05} as well as Meden and
Marquard~\cite{MED06a}.

The effective potential for this problem corresponds to
Eq.~(\ref{method-Uform}) and the flow equations for the running
couplings are given by Eq.~(\ref{method-2com-result}). It is very
easy to solve these equations numerically. In
Fig.~\ref{den-pol-double-dot} we show results for the parameter set
given in the figure caption. This parameter set with the rather
large Coulomb interaction $U/\Gamma=8$ and $t_{12}=0$ has been taken
from Ref.~\onlinecite{MED06a}. (For convenience, here and in the
following all energies will be given in units of $\Gamma=\sum_{il}
\Gamma_i^l$.) The figure shows Coulomb correlation induced
resonances close to $V_g=0$. The dashed curves agree with the
results of Ref.~\onlinecite{MED06a}, which were calculated
neglecting the renormalization of the Coulomb interactions. The full
curves include this renormalization, and the figure clearly shows
its importance (see also Ref.~\onlinecite{KAR07}). We would like to
emphasize, that within the static approximation, there are no higher
order vertex functions to be considered, since the set of equations
(\ref{method-2com-result}) is complete. It is therefore possible to
apply these equations at even larger Coulomb interactions. One finds
that with increasing Coulomb interactions the two outer Coulomb
blockade peaks seen in Fig.~\ref{den-pol-double-dot} shift further
outwards and the exponentially sharp correlation induced peaks in
the center shift exponentially close to zero. The dot level
occupancies displayed in Fig.~\ref{den-pol-double-dot} show the
characteristic charge oscillations discussed in detail in
Refs.~\onlinecite{KOEN05} and~\onlinecite{SIN05}.

\subsection{Side coupled double dot}

\begin{figure}
\unitlength1cm
\begin{picture}(18,3)(0,0)
 \put(2,0)    {\includegraphics[width=5cm]{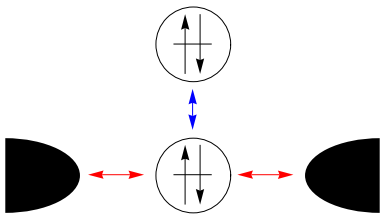}}
\end{picture}
\caption{\label{side-coupled}\it (color online) Side-coupled quantum dot device.}
\end{figure}

Now we consider a side coupled double dot with spin as depicted in
Fig.~\ref{side-coupled}. Only one dot is coupled to the leads
directly. For this system, the set of coupled flow equations is much
more complicated than for the polarized double dot briefly reviewed
above, and higher order equations $m>2$ play a role. In fact,
assuming spin symmetric interactions, we have to solve a set of 26
coupled equations as was pointed out before. We will investigate two
physically different cases: large interdot hopping $t_{12}>\Gamma$
and small interdot hopping $t_{12}<\Gamma$. In both cases we study
the influence of intra-dot Coulomb interactions $U$, which are
chosen to be equal on both quantum dots. Inter-dot Coulomb
interactions are set initially to zero, but they evolve to finite
values during the renormalization run. In fact, all 26 running
couplings (vertex functions) considered here renormalize to non-zero
values in certain gate voltage ranges, some of them attaining
extremely large values. Side coupled double dots have been
investigated by Cornalia and Grempel~\cite{COR05} and Zitko and
Bonca~\cite{ZIT06} using the NRG as well as  Karrasch, Enss and
Meden~\cite{KAR06} using the fRG truncated at $m=2$.
\begin{figure*}
\unitlength1cm
\begin{picture}(18,5)(0,0)
 \put(0,0.0) {\includegraphics[width=4.8cm]{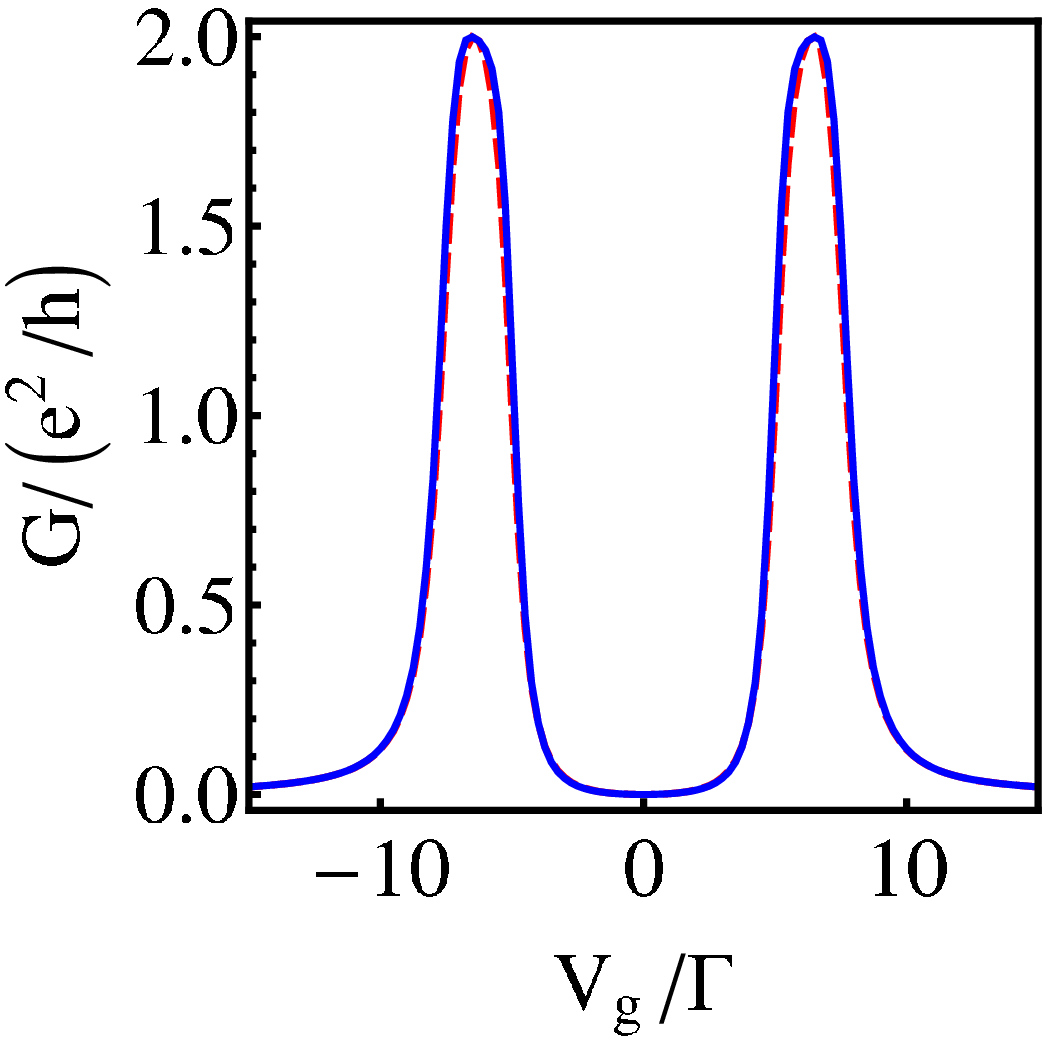}}
 \put(4.2, 4.2) {(a)}
 \put(5.3,0)    {\includegraphics[width=4.8cm]{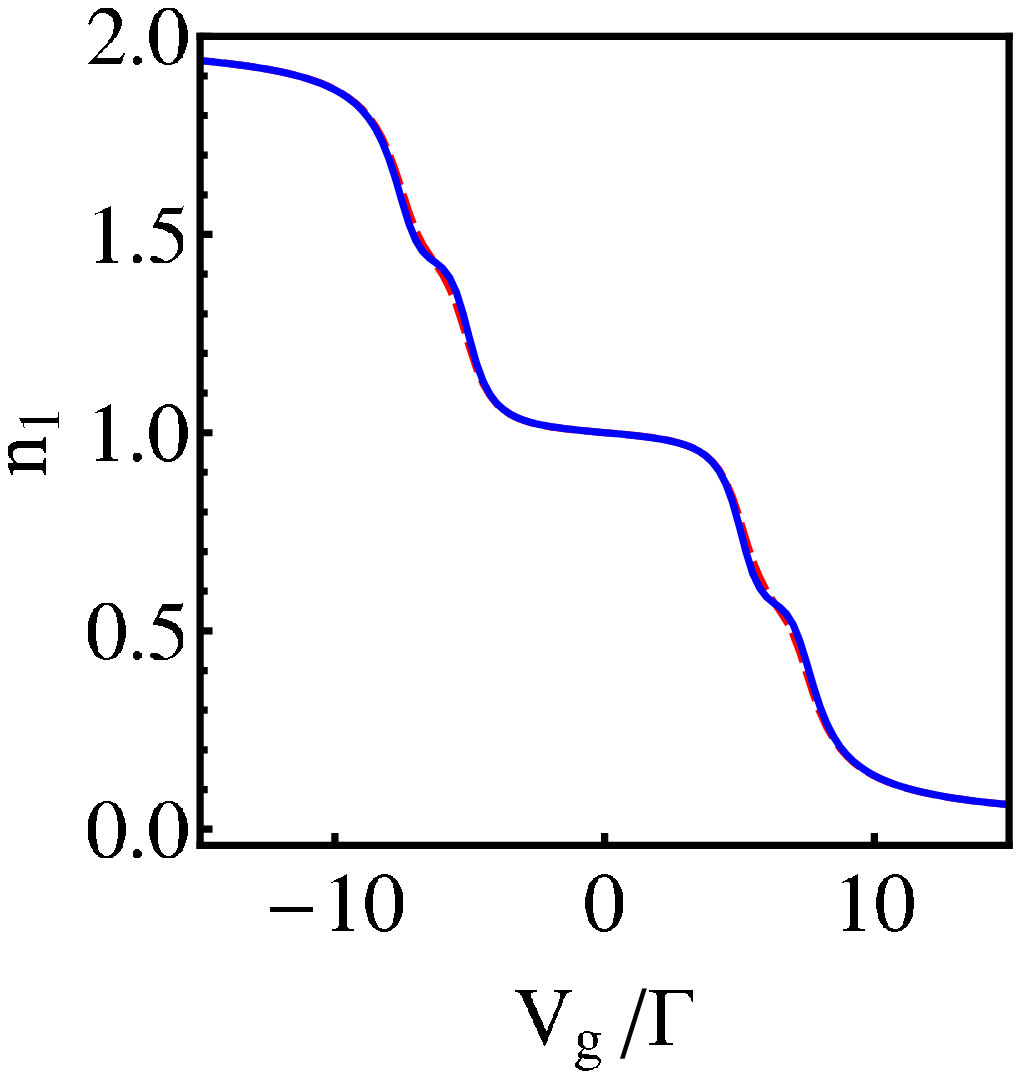}}
 \put(9.5, 4.2) {(b)}
 \put(10.6,0.0){\includegraphics[width=4.8cm]{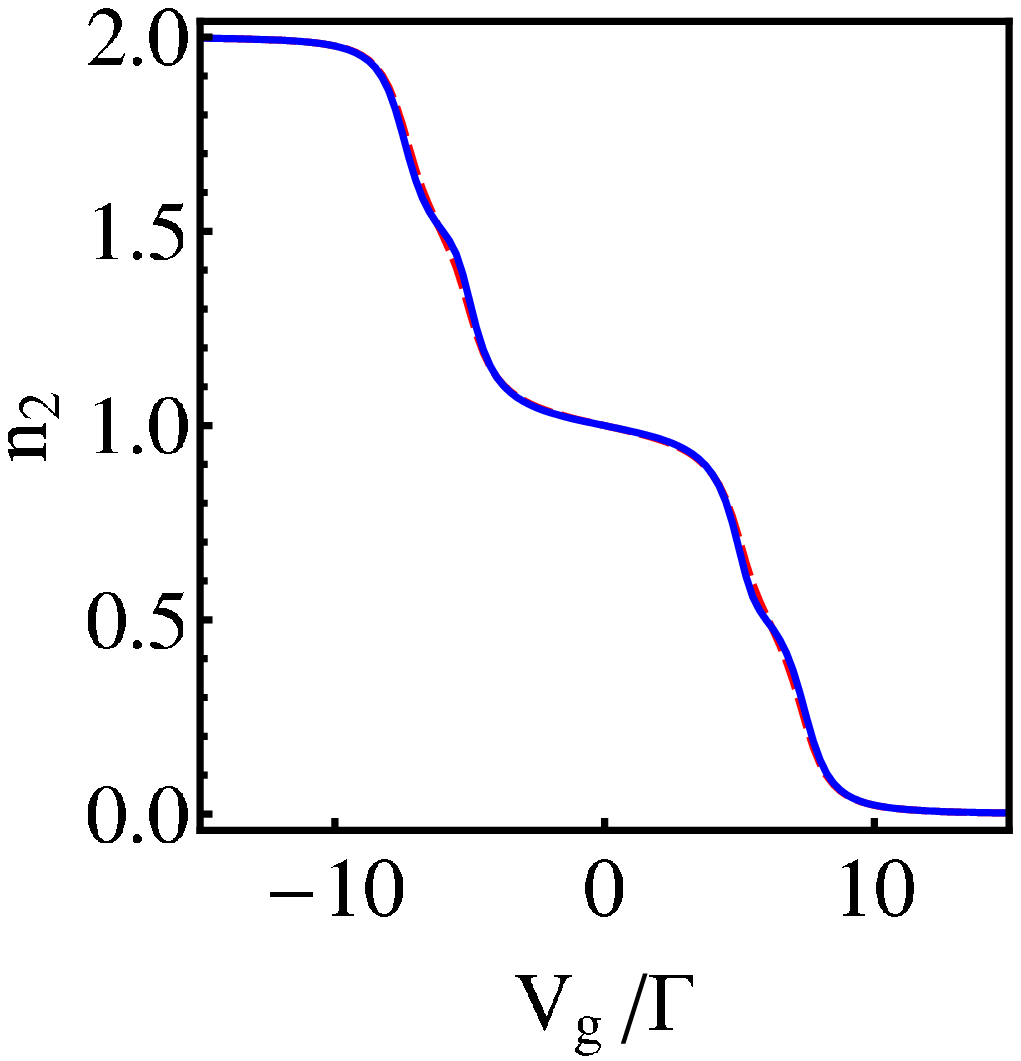}}
 \put(14.8, 4.2) {(c)}
\end{picture}
\caption{\label{side-dot-U8}\it (color online) Side coupled double
dot with large interdot hopping: $t_{12}/\Gamma=4$, $U/\Gamma=8$,
$\Gamma_1^L=\Gamma_1^R=\Gamma/2$, full  lines: truncation order
$m=4$, dashed lines: truncation order $m=2$, (a) conductance
$\mathcal{G}$, (b) occupancy of directly coupled dot $n_1$, (c)
occupancy of side coupled dot $n_2$.}
\end{figure*}

We start with a discussion of dot systems with large interdot
hopping described by the parameter set $t_{12}/\Gamma=4$,
$U/\Gamma=8$, and $\epsilon_{1\sigma}=\epsilon_{2\sigma}=0$ and
other parameters given in the caption of Fig.~\ref{side-dot-U8}. The
same parameter set was discussed in Refs.~\onlinecite{KAR06}
and~\onlinecite{COR05}. Rough estimates discussed in
Ref.~\onlinecite{COR05} suggest that one should observe conductance
peaks located at $V_g\approx\pm (U/2+t_{12})$ with a width given by
$U/2$. This is indeed seen in Fig.~\ref{side-dot-U8}(a). In this
figure, we compare calculations truncated at different flow orders.
The dashed curves show calculations truncated at $m=2$ as in
Ref.~\onlinecite{KAR06}. The full curves include all orders up to
$m=4$, which is the maximum order possible in this system. We
observe that the inclusion of the higher orders has the effect of
somewhat increasing the width of the peaks in accordance with the
NRG calculations presented in Ref.~\onlinecite{COR05}. The dot
occupation numbers [Fig.~\ref{side-dot-U8}(b)and (c)] show slightly
more pronounced shoulders if higher order effects $m>2$ are included
within the gate voltage ranges of peak conductance.

\begin{figure*}
\unitlength1cm
\begin{picture}(18,5)(0,0)
 \put(0.,0.0)    {\includegraphics[width=4.8cm]{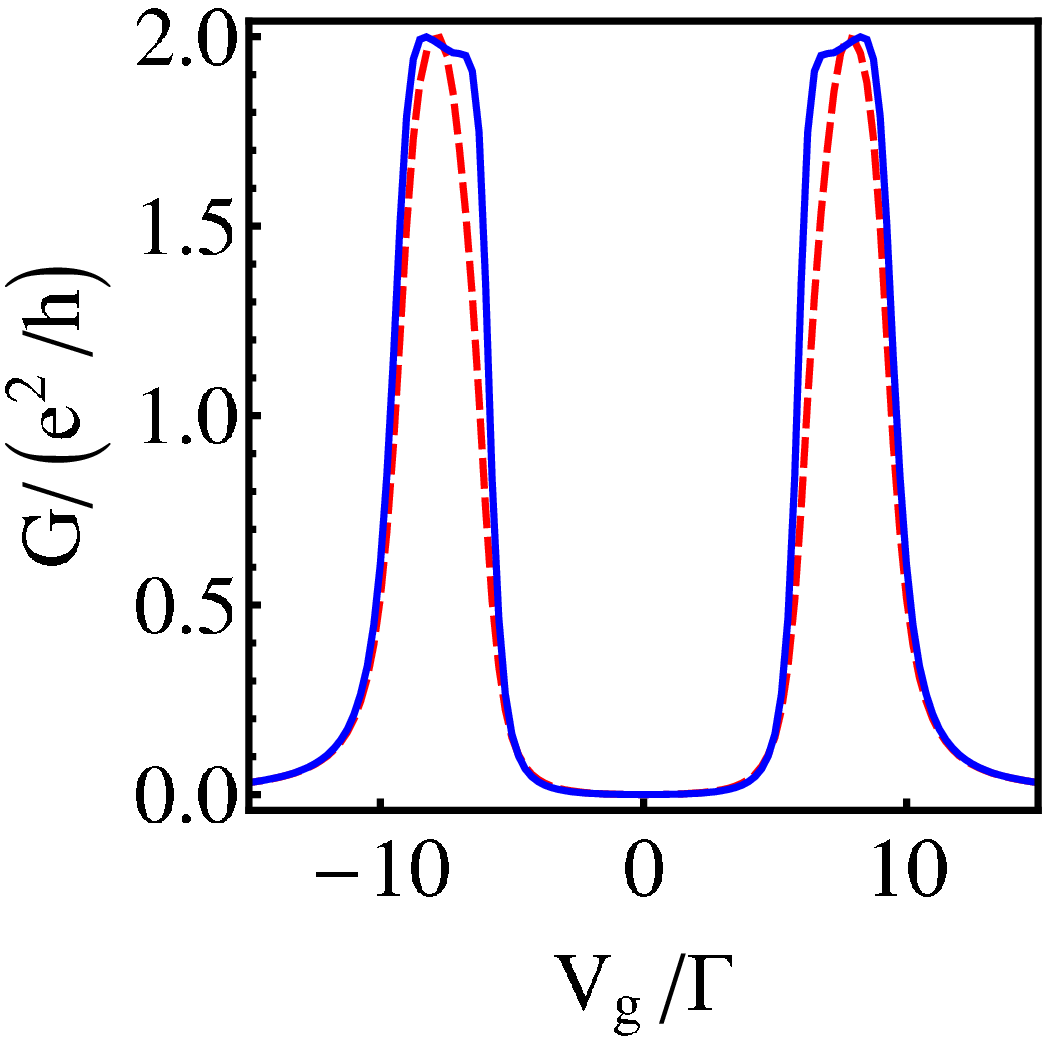}}
 \put(4.2, 4.2)  {(a)}
 \put(5.3,0)     {\includegraphics[width=4.8cm]{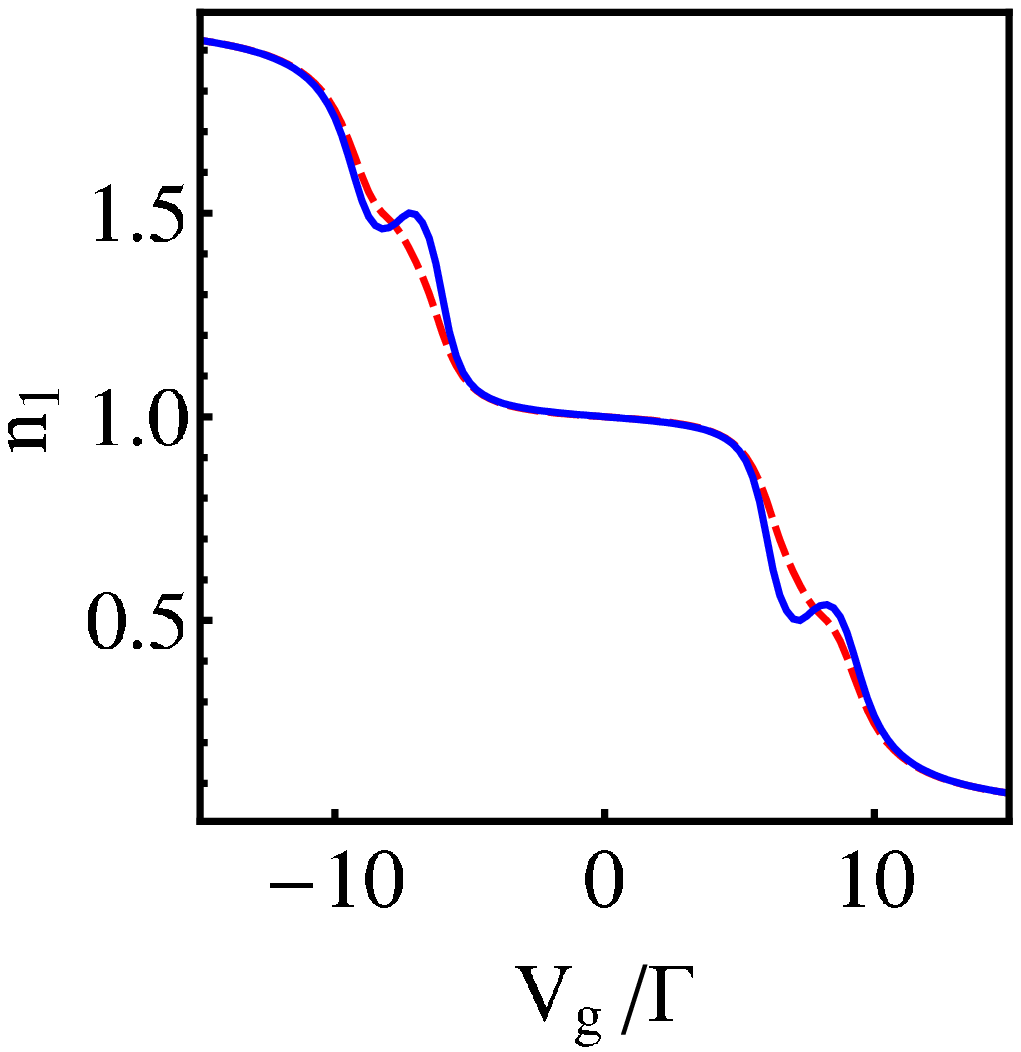}}
 \put(9.5, 4.2)  {(b)}
 \put(10.6,0.0)  {\includegraphics[width=4.8cm]{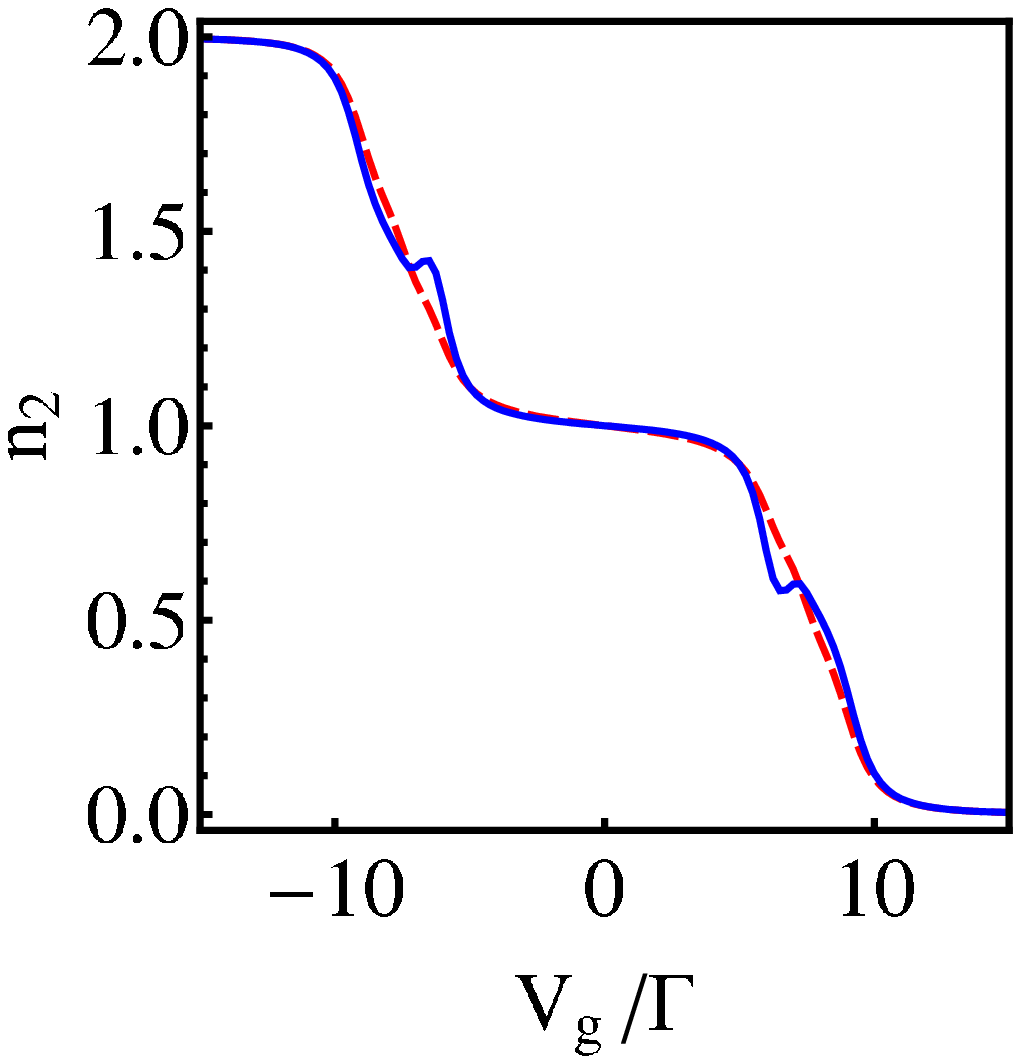}}
 \put(14.8, 4.2) {(c)}
\end{picture}
\caption{\label{side-dot-U12}\it (color online) Side coupled double
dot with large interdot hopping: $t_{12}/\Gamma=4$, $U/\Gamma=12$.
Other parameters and notation as in Fig.~\ref{side-dot-U8}}.
\end{figure*}

Now we increase the intra-dot Coulomb interaction to $U/\Gamma=12$.
According to the discussion presented in Ref.~\onlinecite{ZIT06},
the peaks should become a more box-like in shape. This is indeed
seen in Fig.~\ref{side-dot-U12}(a), but only if the higher order
contributions $m>2$ are included. A calculation  truncating the set
of equations at $m=2$ shows conductance peaks with a Lorentzian
shape in disagreement with NRG calculations~\cite{ZIT06}. The
conductance peaks shown in Fig.~\ref{side-dot-U12}(a) are not
completely flat as suggested by the NRG calculations. This probably
points to effects beyond the static approximation employed here. The
occupancies of the two dots are shown in Fig.~\ref{side-dot-U12}(b)
and (c). Here the effect of higher order contributions is quite
important as well. The single occupancies $n_i$ even show a
non-monotonic behavior in the gate voltage ranges of peak
conductance. Similarly, the transmission phase develops a shoulder
in these gate voltage ranges which is not seen in calculations
truncated at $m=2$.

If one increases the intra-dot Coulomb interaction further, e.g., to
$U/\Gamma=16$, one observes conductance peaks which are flat, but
within this flat region the renormalization procedure employed here
becomes ill defined. Technically, this can be traced to the fact
that somewhere along the renormalization flow the renormalized self
energies of the indirectly coupled dot vanish. At this point the
body of the matrix $K$ which was introduced in the previous section
vanishes, {\it all} flow equations develop a singularity, and the
renormalization flow stops. The singular renormalization flow likely
indicates a failure of the static approximation.

\begin{figure*}
\unitlength1cm
\begin{picture}(18,5)(0,0)
 \put(0.,0.0)    {\includegraphics[width=4.8cm]{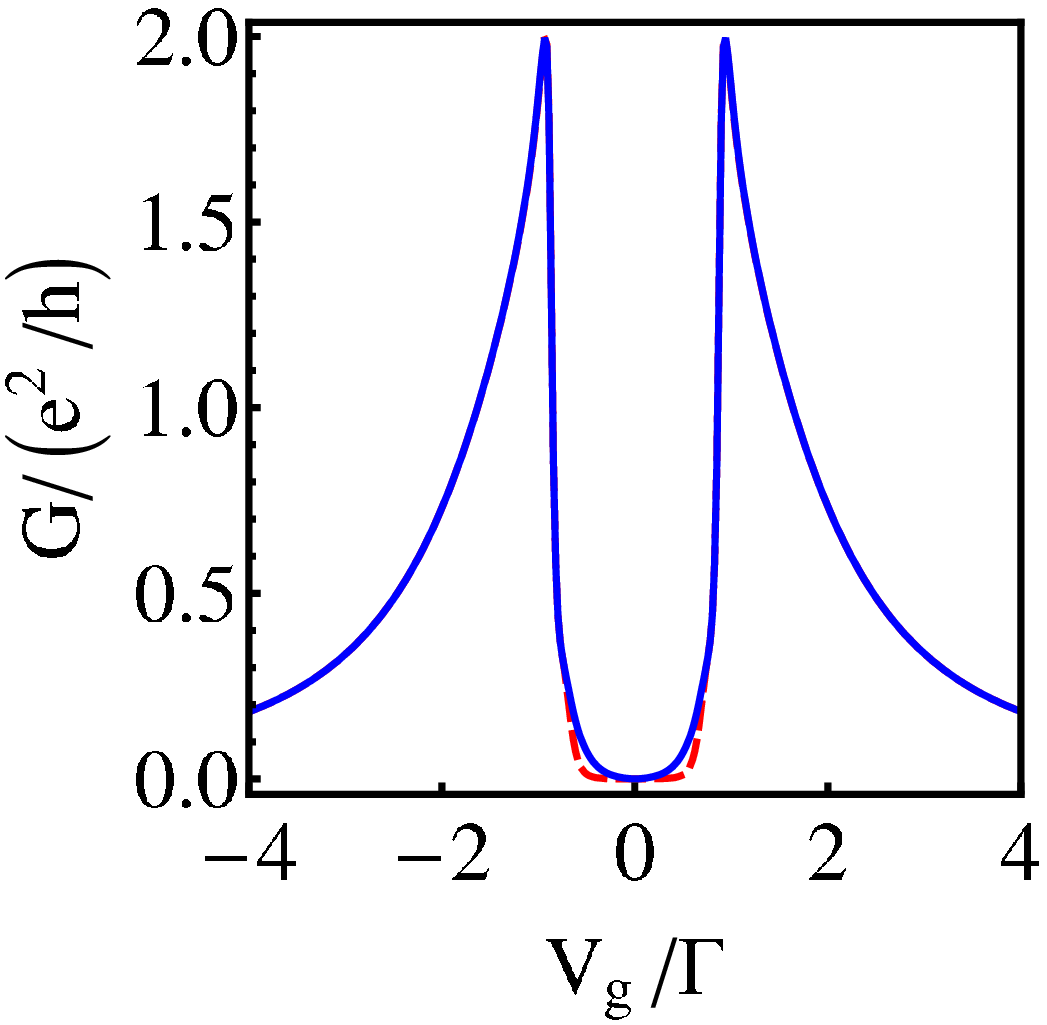}}
 \put(4.1, 4.2)  {(a)}
 \put(5.3,0)     {\includegraphics[width=4.8cm]{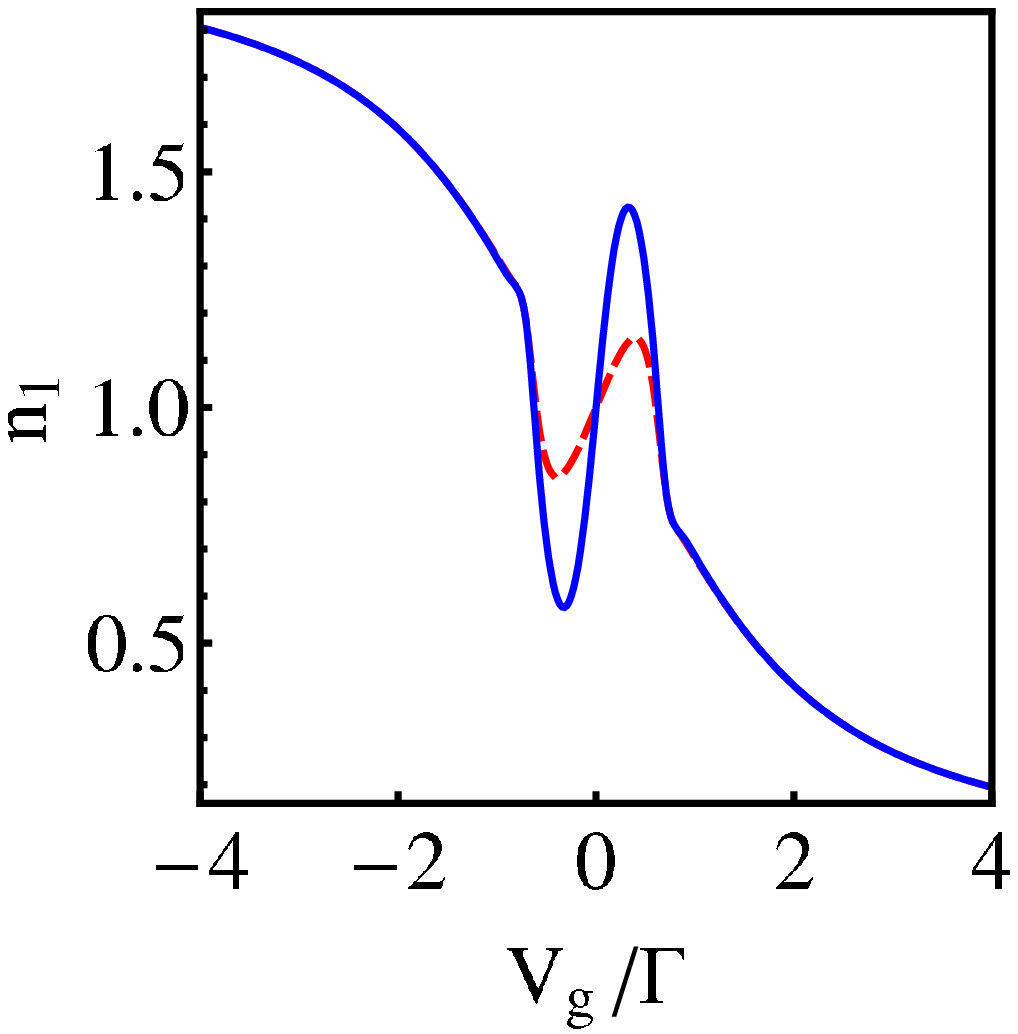}}
 \put(9.4, 4.2)  {(b)}
 \put(10.6,0.0)  {\includegraphics[width=4.8cm]{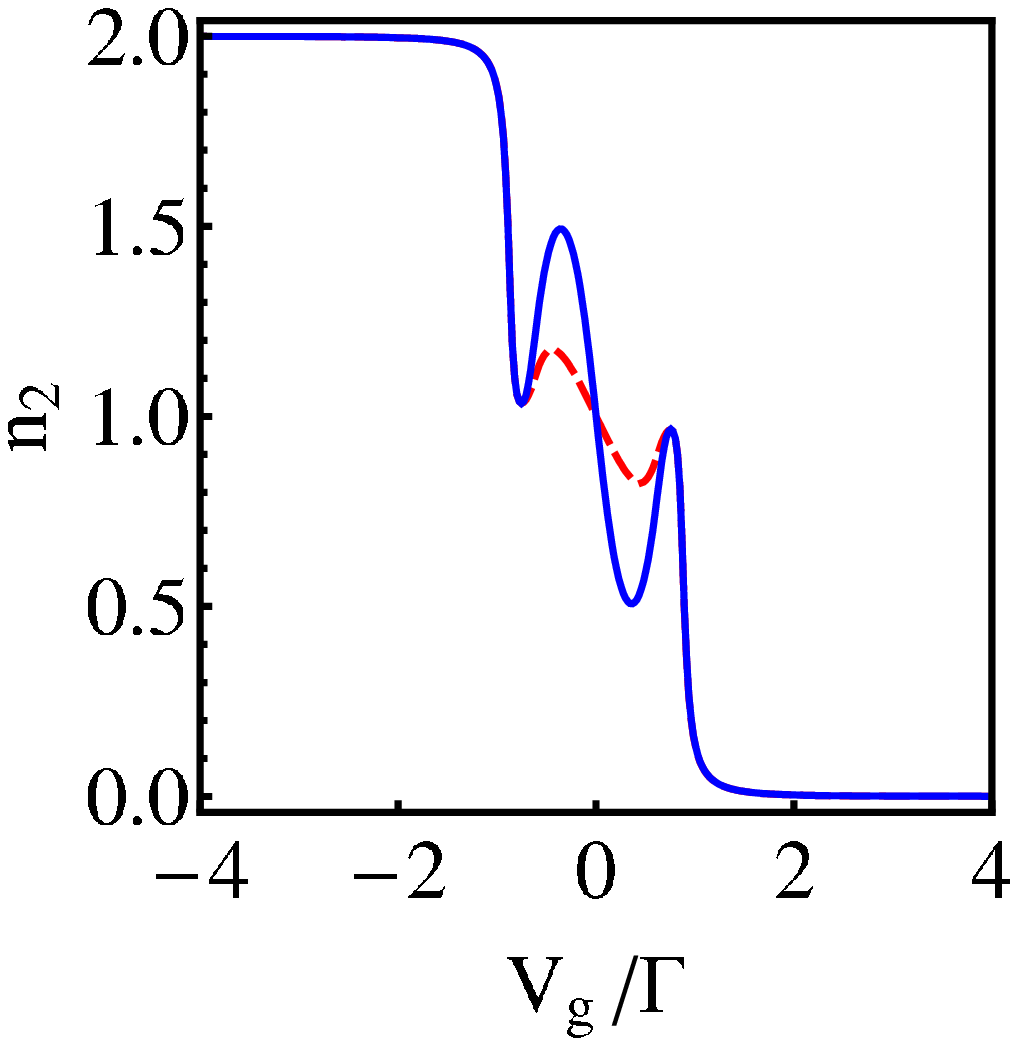}}
 \put(14.7, 4.2) {(c)}
\end{picture}
\caption{\label{side-dot-U2}\it (color online) Side coupled double
dot with small interdot hopping $t_{12}/\Gamma=.2$, $U/\Gamma=2$.
Other parameters and notation as in Fig.~\ref{side-dot-U8}}.
\end{figure*}

Let us now discuss a physical situation with small inter-dot
hopping: Fig.~\ref{side-dot-U2} shows a calculation using the
parameters $t_{12}/\Gamma=.2$ and $U/\Gamma=2$,
$\epsilon_{1\sigma}=\epsilon_{2\sigma}=0$. This parameter set was
also investigated in Refs.~\onlinecite{KAR06}
and~\onlinecite{COR05}. As was pointed out by the authors of
Ref.~\onlinecite{KAR06}, the calculation truncated at $m=2$ yields a
conductance minimum, which is too flat in comparison to NRG
calculations~\cite{COR05}. As is obvious from
Fig.~\ref{side-dot-U2}(a) inclusion of higher order effects improves
this situation. However, comparison with the NRG data presented in
Ref.~\onlinecite{COR05} still shows quantitative differences in this
gate voltage region. Since our calculations are not truncated within
the constraints of the static approximation , these differences must
be attributed to effects beyond the static approximation.

\begin{figure*}
\unitlength1cm
\begin{picture}(18,5)(0,0)
 \put(2,0)    {\includegraphics[width=4.8cm]{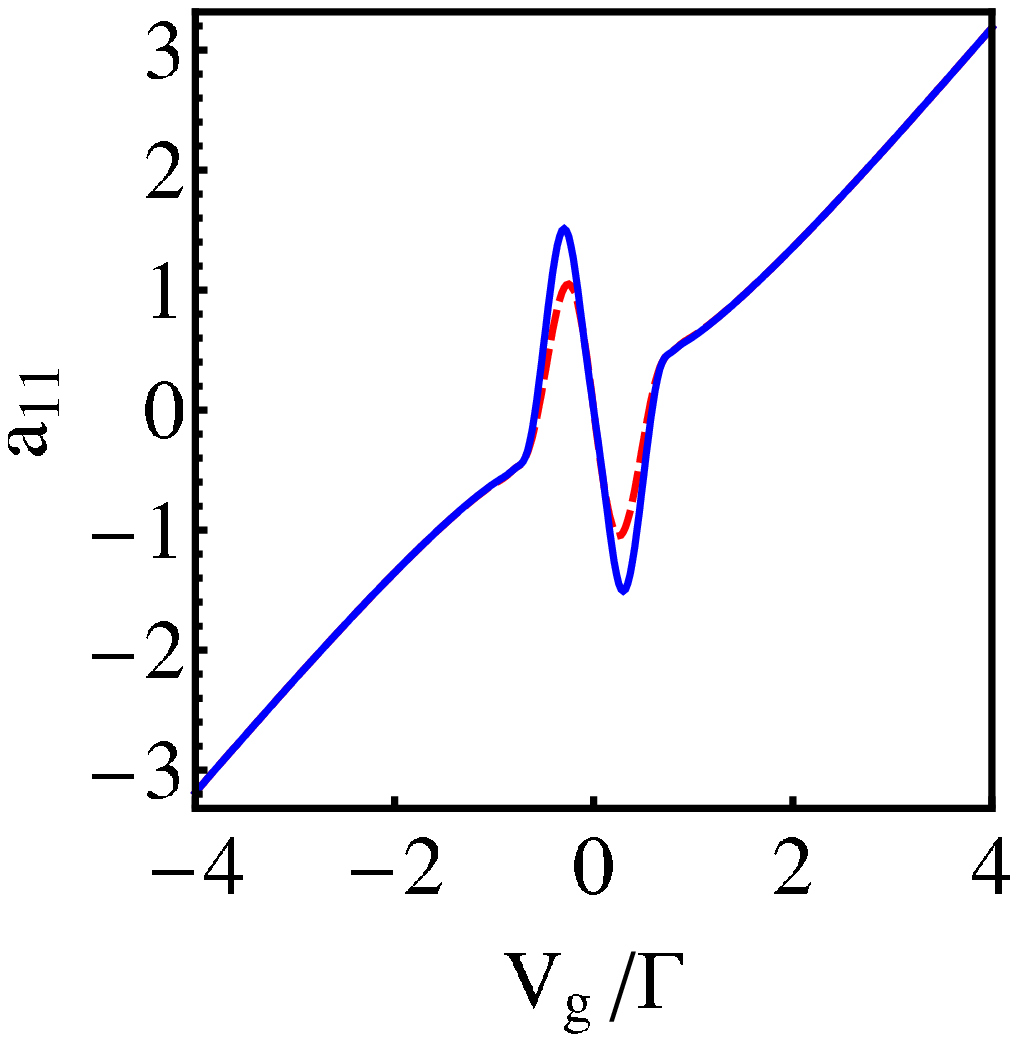}}
 \put(8.0,0.0){\includegraphics[width=4.8cm]{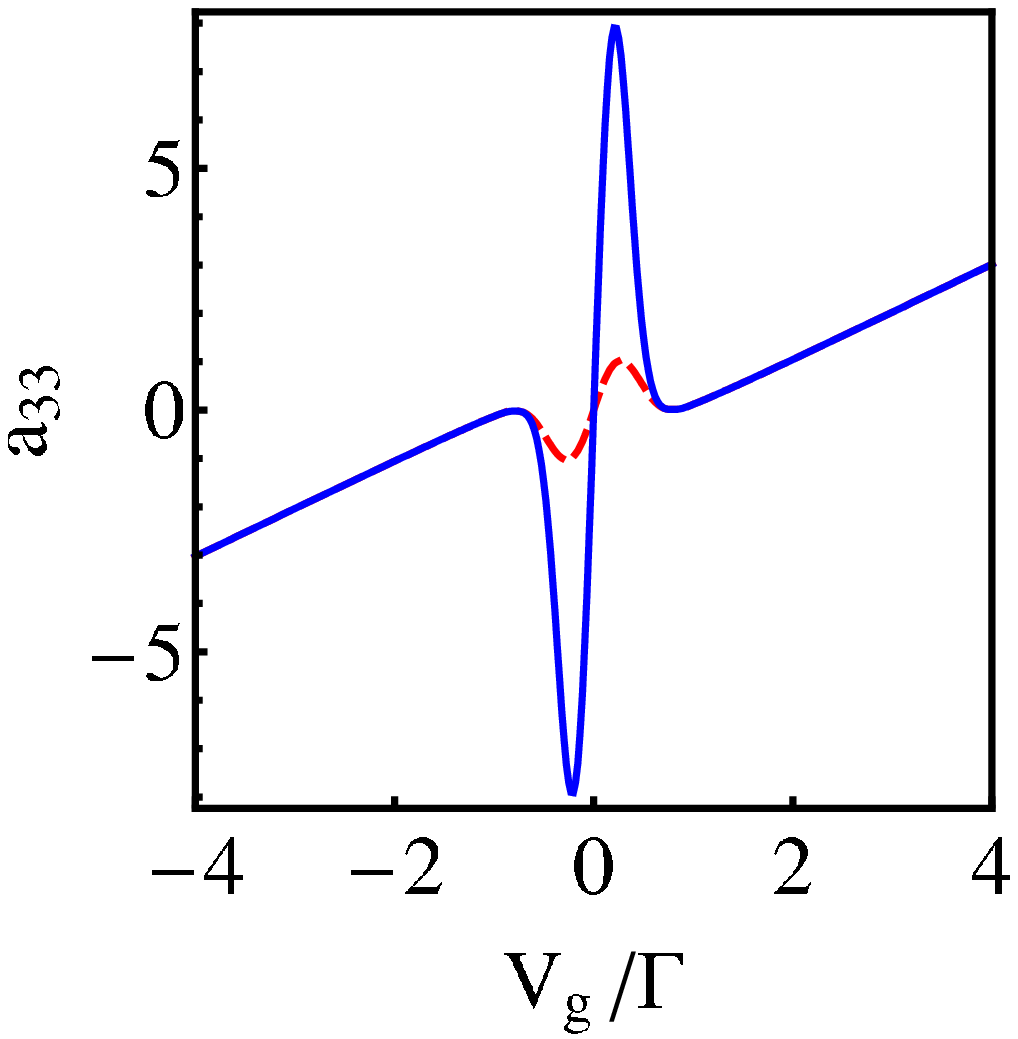}}
\end{picture}
\caption{\label{side-dot-U2-energies}\it  (color online) Energy
eigenvalues of the spin up level in directly coupled dot ($a_{11}$)
and the indirectly coupled dot ($a_{33}$) as a function of the gate
voltage for the parameter set of Fig.~\ref{side-dot-U2}.}
\end{figure*}

Of course, the structure obtained for the conductance and the
occupancies as a function of the gate voltage can be related to the
calculated spectrum of the dot system. In
Fig.~\ref{side-dot-U2-energies} we show the energy level for a spin
up electron on the directly coupled dot given by $a_{11}$ and the
energy level of a spin up electron on the side-coupled dot $a_{33}$
for the parameter set used in Fig.~\ref{side-dot-U2}. Again we
compare calculations truncated at $m=2$ with a full calculation in
the static approximation. Obviously, truncating at $m=2$ yields
significantly too small energy eigenvalues in particular for the
indirectly coupled dot.

If we increase the intra-dot Coulomb interaction from $U/\Gamma=2$
to $U/\Gamma=2.5$, we observe a picture similar to the one shown in
Fig.~\ref{side-dot-U2} with the amplitude of the charge oscillations
on the dots significantly increased. Already for $U/\Gamma=3$, the
set of flow equations develops singular behavior as was discussed
above  for the case with large inter-dot hopping. This again
indicates that a truncation scheme based on a static approximation
fails for large Coulomb interactions .

\subsection{Parallel coupled double dot}

\begin{figure}
\unitlength1cm
\begin{picture}(18,3)(0,0)
 \put(2,0)    {\includegraphics[width=5cm]{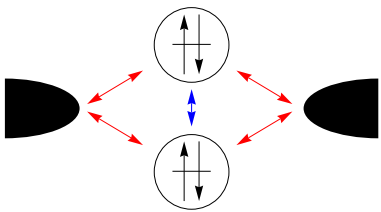}}
\end{picture}
\caption{\label{parallel-coupled}\it  (color online) Parallel coupled quantum dot.}
\end{figure}

Finally we consider a parallel coupled double dot as depicted in
Fig.~\ref{parallel-coupled}. The set of equations to be solved is
the same as for the side-coupled dot.

\begin{figure*}
\unitlength1cm
\begin{picture}(18,5)(0,0)
 \put(0.,0.0)    {\includegraphics[width=4.8cm]{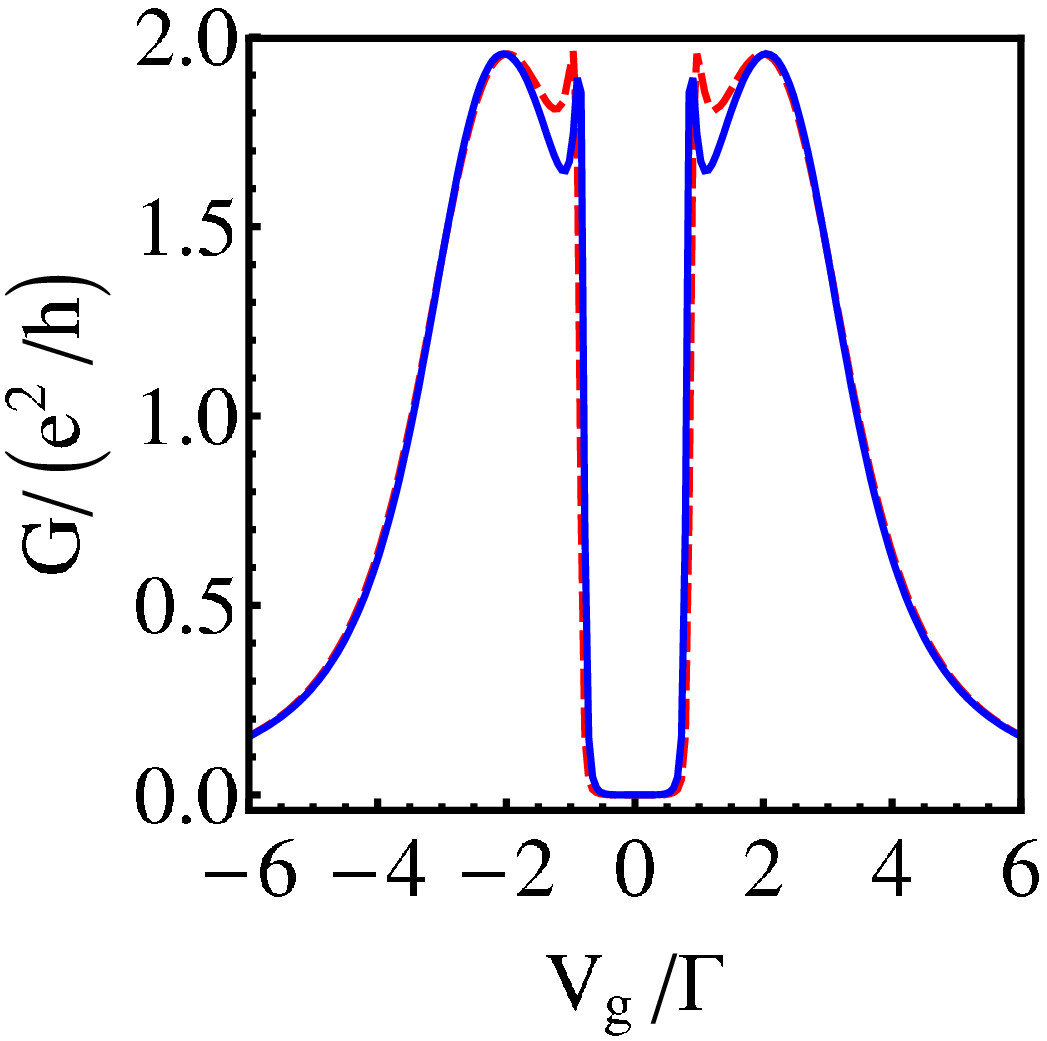}}
 \put(4.1, 4.2)  {(a)}
 \put(5.3,0)     {\includegraphics[width=4.8cm]{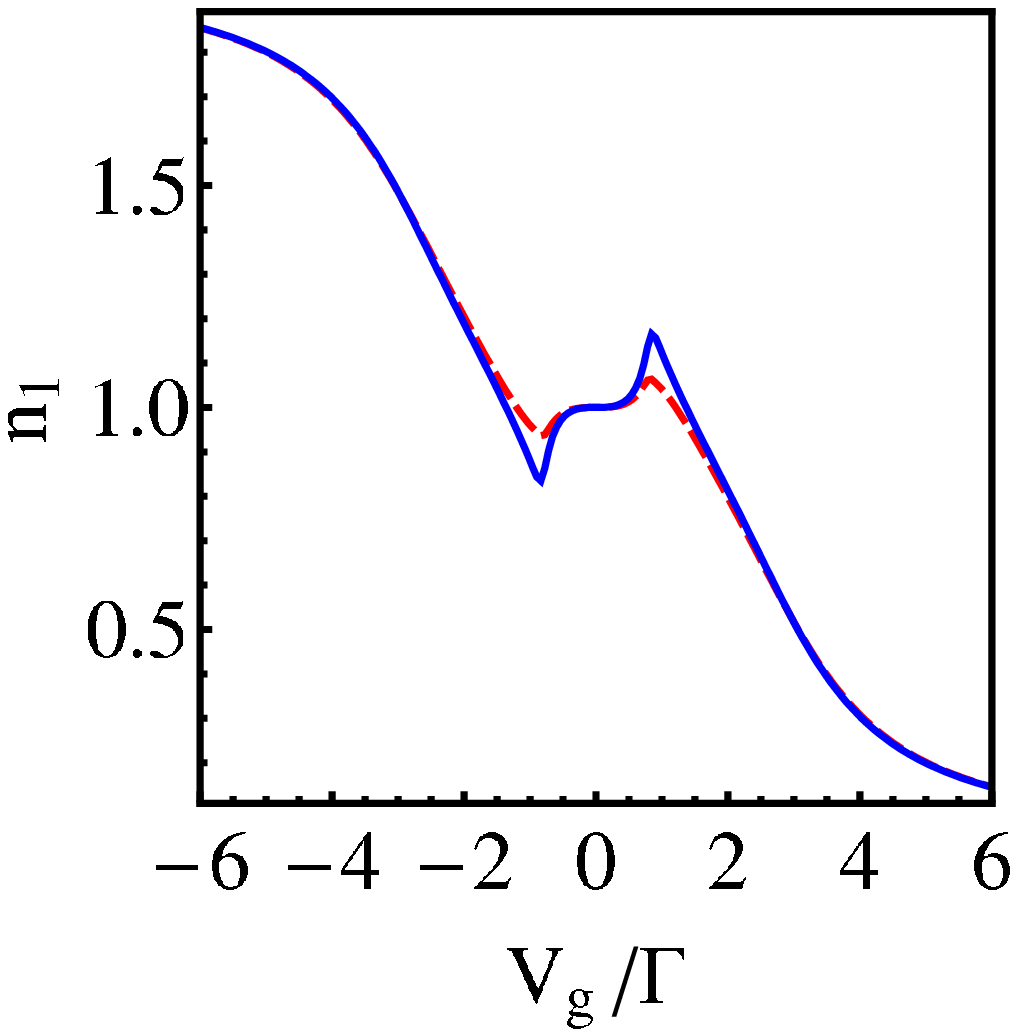}}
 \put(9.4, 4.2)  {(b)}
 \put(10.6,0.0)  {\includegraphics[width=4.8cm]{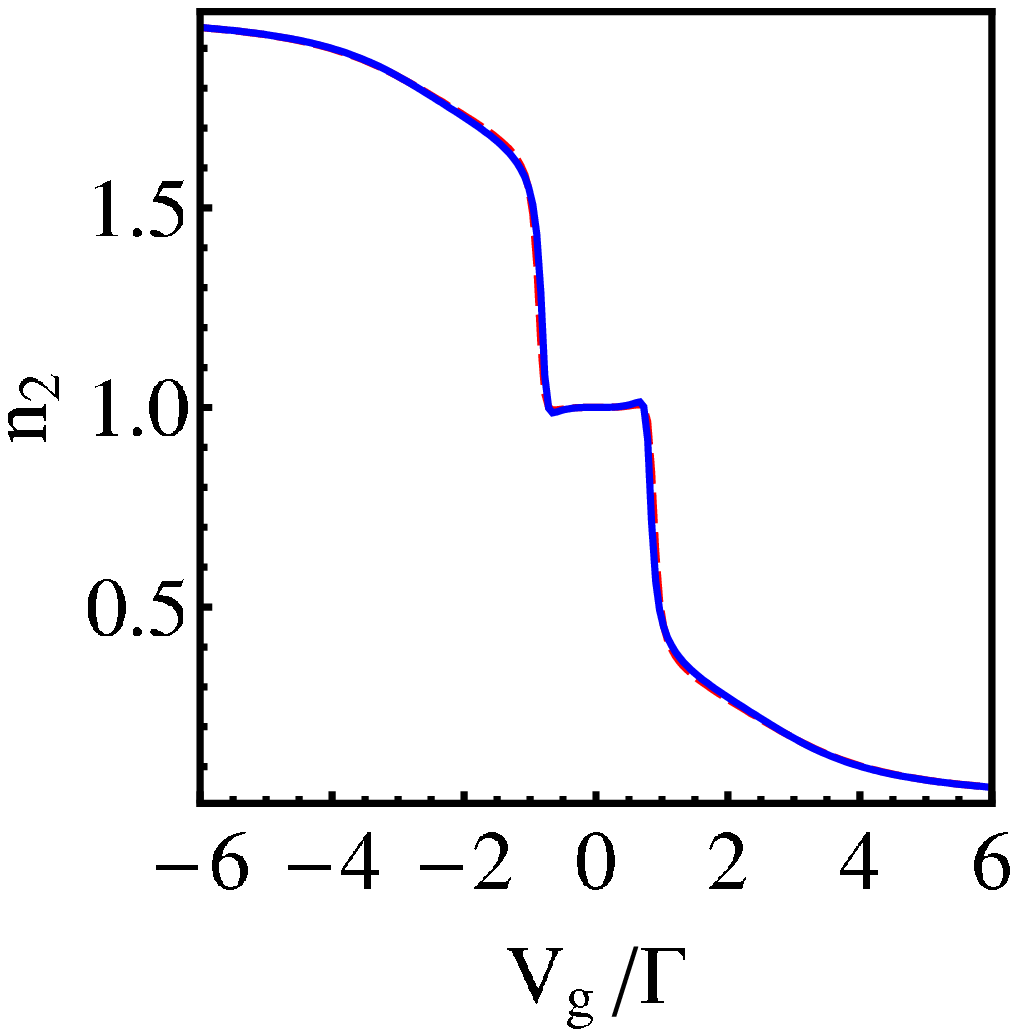}}
 \put(14.7, 4.2) {(c)}
\end{picture}
\caption{\label{para-dot-U2}\it  (color online) Parallel coupled
double dot $t_{12}/\Gamma=0$, $U/\Gamma=2$, $\Gamma_1^L/\Gamma=.5$,
$\Gamma_1^R/\Gamma=.25$, $\Gamma_2^L=0.07$, $\Gamma_2^R=0.18$. Other
notation as in Fig.~\ref{side-dot-U8}}.
\end{figure*}

\begin{figure*}
\unitlength1cm
\begin{picture}(18,5)(0,0)
 \put(0.,0.0)    {\includegraphics[width=4.8cm]{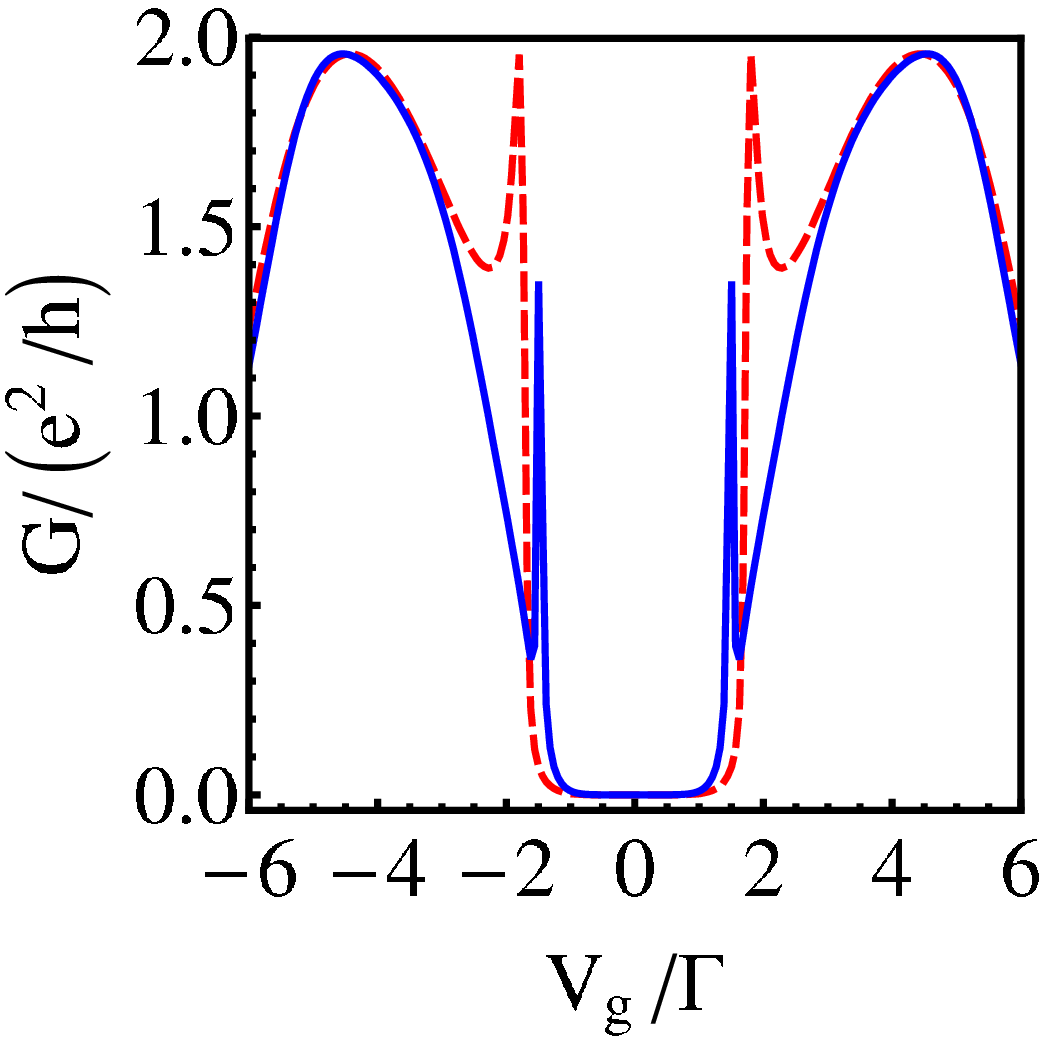}}
 \put(4.1, 1.4)  {(a)}
 \put(5.3,0)     {\includegraphics[width=4.8cm]{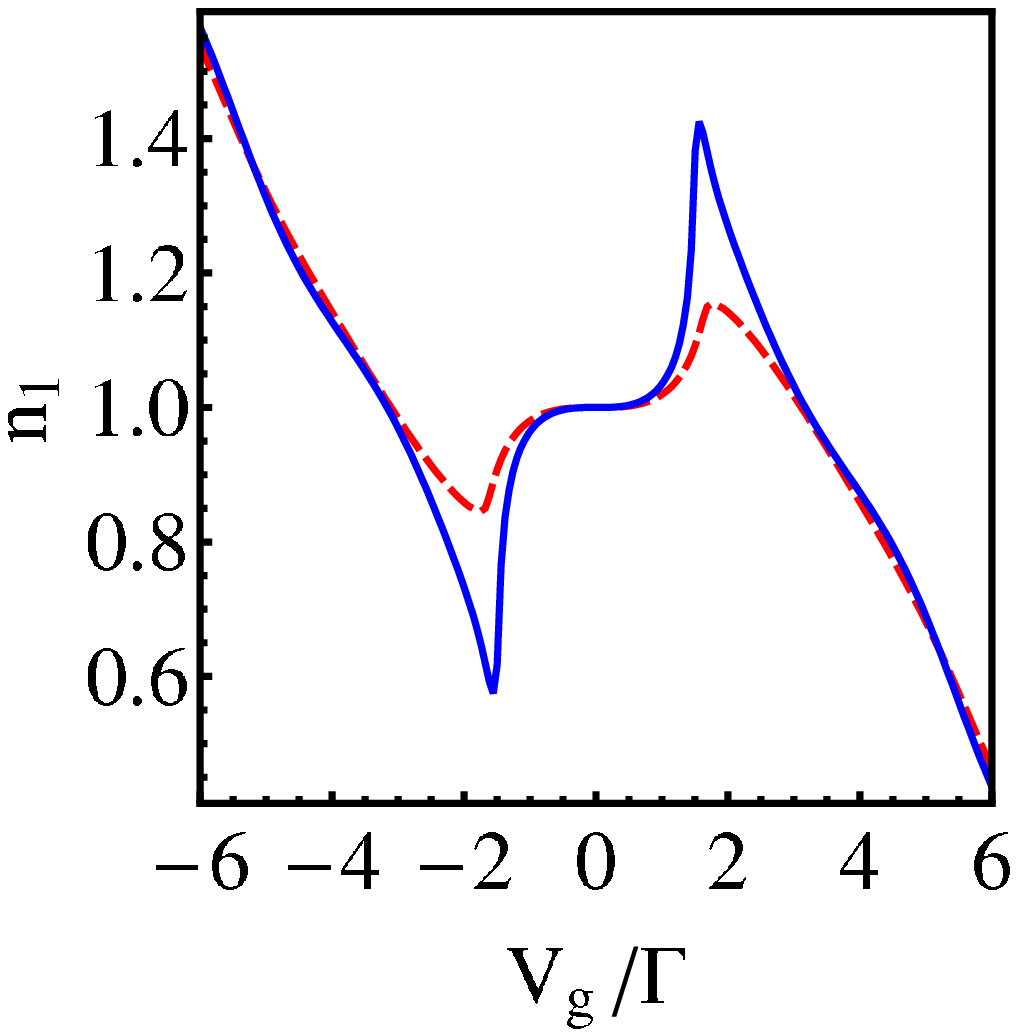}}
 \put(9.4, 4.2)  {(b)}
 \put(10.6,0.0)  {\includegraphics[width=4.8cm]{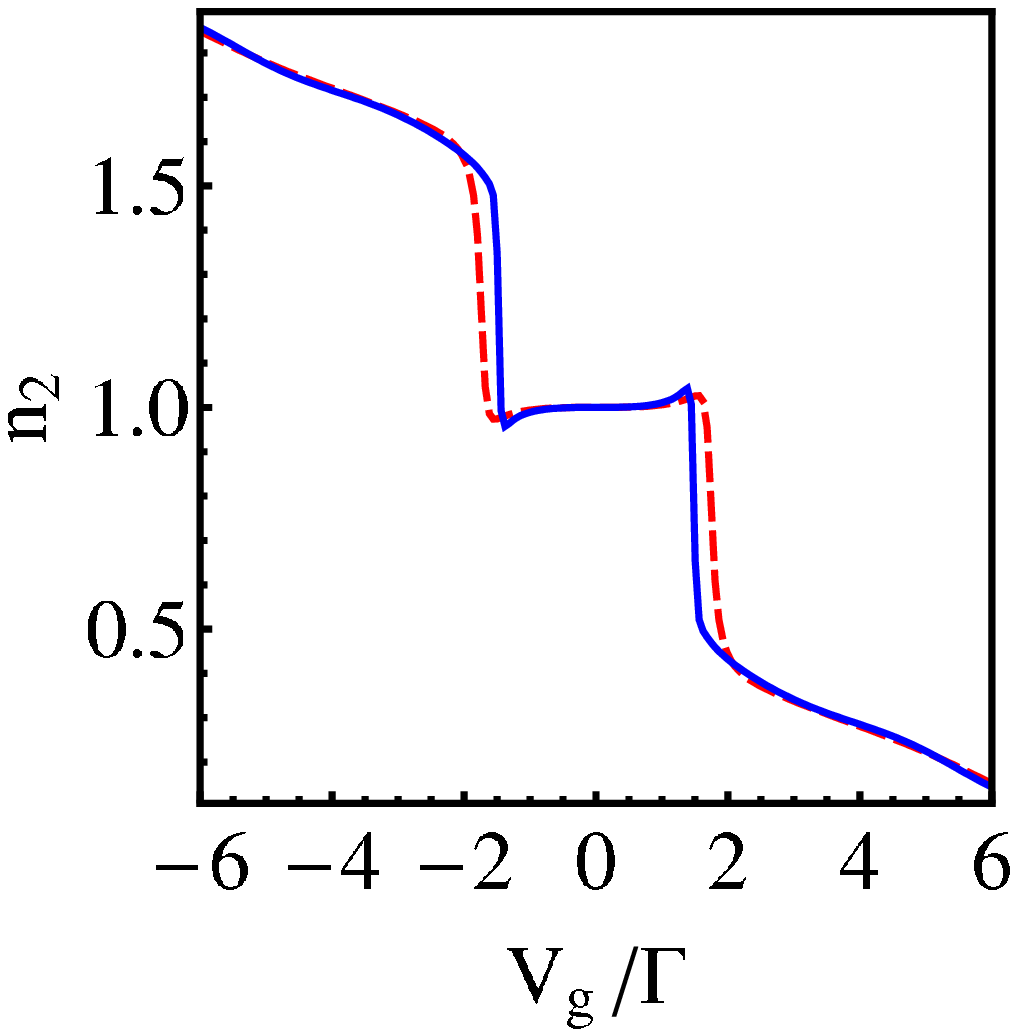}}
 \put(14.7, 4.2) {(c)}
\end{picture}
\caption{\label{para-dot-U4}\it  (color online) Parallel coupled
double dot $t_{12}/\Gamma=0$, $U/\Gamma=4$, $\Gamma_1^L/\Gamma=.5$,
$\Gamma_1^R/\Gamma=.25$, $\Gamma_2^L=0.07$, $\Gamma_2^R=0.18$. Other
notation as in Fig.~\ref{side-dot-U8}}.
\end{figure*}

We study a parameter set where all $t_i^l$ are chosen positive with
the tunneling strengths $\Gamma_i^l$  given numerically in the
caption of Fig.~\ref{para-dot-U2}.  We choose
$U_{jj^\prime}^{\sigma\sigma^\prime}/\Gamma=2$ for all intra-dot and
inter-dot interaction matrix elements and $t_{12}=0$. This parameter
set has also been investigated in Ref.~\onlinecite{KAR06}. Due to
the large Coulomb interaction, correlation induced resonances are
observed similar to the spin-polarized system briefly reviewed in
section~\ref{polDD}. However, as is seen in Fig.~\ref{para-dot-U2}
calculations truncated at $m=2$ show significant quantitative
differences in comparison to the full calculation up to $m=4$: The
resonances are much more pronounced and the valley between these
resonances is somewhat narrower.
This trend continues, if we increase the Coulomb interactions to
$U_{jj^\prime}^{\sigma\sigma^\prime}/\Gamma=4$. Results are shown in
Fig.~\ref{para-dot-U4}. The amplitude of the charge oscillations of
the dot with the stronger tunnel coupling to the leads is
significantly underestimated in a calculation neglecting higher
order contributions.

Of course, the question arises, to what extend the static
approximation is still valid in this parameter region. In order to
answer this question calculations beyond the static approximation
must be performed. Such calculations are presently under way.

\section{Conclusions}

The functional renormalization group has been proposed as a useful
new tool for the investigation of interacting mesoscopic
systems~\cite{KAR06, MED06a}. The advantage of the method in
comparison to NRG or DMRG calculations appears to be its rather
moderate numerical cost. However, this advantage is due to some
rather drastic approximations, which are not easily controlled. In
this paper we carry out the first steps that are needed to go beyond
the standard set of approximations  usually employed to solve the
fermionic fRG in practice. Still within the constraints of a static
approximation we develop a method which  systematically generates
the complete set of ordinary differential equations corresponding to
the fRG. However, the method proposed here is not limited to the
static approximation. It can be extended straightforwardly to
include non-static effects e.g. by including a wave function
renormalization in the kinetic energy term of the action.

We demonstrate by means of a number of calculations of transport
properties of quantum dot systems that inclusion of higher order
equations definitely improves fRG calculations for large Coulomb
interactions. This is judged by a comparison to NRG results.
However, if one increases Coulomb interactions beyond some critical
value, the static approximation fails as is signalled by
singularities developing during the renormalization flow. In order
to overcome such problems and to further improve the agreement
between NRG and fRG calculations it is necessary to go beyond the
static approximation. Work in this direction is presently under way.

\begin{acknowledgments}
M.W. would like to thank Volker Meden for many useful discussions
and Christoph Karrasch for providing detailed results of his
calculations. We thank Marcel Reginatto for suggestions in the early
stages of this work and a careful reading of the manuscript.
\end{acknowledgments}

\appendix

\section{Integral over the Matsubara frequencies}\label{app1}

In order to perform the rather singular integral over the Matsubara
frequencies $\omega$ for the hard cutoff regulator
Eq.~(\ref{method-regulator}) we need the formula~\cite{MOR94}
\begin{equation}\label{theta formula}
\delta(x-y)f(\theta(x-y)) \rightarrow \delta(x-y) \int_{0}^{1} ds
f(s).
\end{equation}
The $\omega$-integration leading from Eq.~(\ref{G-renorm}) to Eq.~(\ref{method-floweqa}) is then performed as follows
\begin{widetext}
\begin{eqnarray}
& &{\rm Tr} \int_{-\infty}^\infty \frac{{\rm d}\omega}{2\pi} \left(G^{-1}(i\omega)+Ck\theta(k^2-\omega^2)\mathbb{E} \right)^{-1}                            \frac{\partial}{\partial k}Ck\theta(k^2-\omega^2)\nonumber\\
& &\;\;\;\;=\frac{2N}{\pi}+
{\rm Tr} \int_{-\infty}^\infty \frac{{\rm d}\omega}{2\pi}
                           \left(G^{-1}(i\omega)+Ck\theta(k^2-\omega^2)\mathbb{E} \right)^{-1}
                            {2Ck^2\delta(k^2-\omega^2)}
\end{eqnarray}
\end{widetext}
with $G^{-1}(i\omega)=(i\omega+u_k^{(2)})\mathbb{E}^\prime$. The
quantity $u_k^{(2)}$ represents the body of $U_k^{(2)}$. In the
first term we already performed the limit $C\rightarrow\infty$. The
second term is now evaluated with the help of Eq.~(\ref{theta
formula})
\begin{eqnarray}
& &{\rm Tr}\int_{-\infty}^\infty \frac{{\rm d}\omega}{2\pi}
\left((G^{-1}(i\omega)+Ck\theta(k^2-\omega^2)\mathbb{E}\right)^{-1}\nonumber\\
& &\;\;\;\; \;\;\;\; \;\;\;\;  Ck\left(\delta(k-\omega))+\delta(k+\omega)\right)\nonumber\\
& &\;\;\; =\frac{1}{2\pi}\sum_{\lambda=\pm k}{\rm Tr}\int_0^C {\rm d}s^\prime
 \left( \frac{1}{k}G^{-1}(i\lambda)+s^\prime\mathbb{E}\right)^{-1}
                           \nonumber\\
& &\;\;\;\;= -\frac{1}{2\pi}\sum_{\lambda=\pm k}{\rm Tr} \log\left(\frac{1}{k}
 G^{-1}(i\lambda)\right)+ O(\log C)\nonumber
\end{eqnarray}
Therefore we obtain the equation
\begin{eqnarray}
 a_0^\prime
% &=&\frac{1}{4\pi}{\rm Tr} \log\left(\frac{1}{k}
%G^{-1}(i\lambda)\right)+(k \leftrightarrow -k)\nonumber\\
 &=&\frac{1}{4\pi} \sum_{\lambda=\pm k}\log {\rm det}
 \left(\frac{1}{k}
G^{-1}(i\lambda)\right)\nonumber
\end{eqnarray}
The infinite constant gets absorbed into a suitably modified initial condition.

With similar steps as above one obtains the result given in Eq.~(\ref{method-floweqb}). Here it is important to
use the fact that the factors under the trace may be permuted cyclically.

%\bibliography{../../Renormierung}

\end{document}